\documentclass[a4paper,11pt]{article}
\usepackage{amssymb}
\usepackage{graphicx}
\usepackage{float}
\usepackage{amsmath} 
\usepackage{hyperref}
\hypersetup{colorlinks,allcolors=blue}
\usepackage{array}

\usepackage[T1]{fontenc}

\usepackage[rightcaption]{sidecap}
\graphicspath{ {images/} }

\usepackage{subcaption}

\usepackage{jinstpub} % for details on the use of the package, please see the JINST-author-manual

\usepackage[nottoc]{tocbibind}

\setcitestyle{numbers}
\setcitestyle{square}
\setcitestyle{comma}

\title{\boldmath Mechanical design concept version 2.0 for the miniBeBe subsystem of the Multi-Purpose Detector at the Nuclotron-based Ion Collider fAcility of the Joint Institute for Nuclear Research}

\author[a,b]{M. Herrera}
\author[c]{M.E. Pati\~no}
\author[c]{Mauricio Alvarado}
\author[b]{Ivonne Maldonado}
\author[b]{Denis Andreev}
\author[c]{Alejandro Ayala}
\author[c]{Wolfgang Bietenholz}
\author[b]{César Ceballos}
\author[c]{Eleazar Cuáutle}
\author[d]{Isabel Domínguez}
\author[e]{L. A. Hernández}
\author[f]{Israel Luna}
\author[b]{Tuyana Lygdenova}
\author[f]{Pablo Martínez-Torres}
\author[f]{Alfredo Raya}
\author[g]{Ulises Sáenz-Trujillo}
\author[a]{M.E. Tejeda-Yeomans}
\author[g]{Galileo Tinoco-Santillán}

\affiliation[a]{Facultad de Ciencias-CUICBAS, Universidad de Colima, Bernal Díaz del Castillo No. 340, Col. Villas San Sebastián, Colima, 28045, Colima, Mexico}

\affiliation[b]{Joint Institute for Nuclear Research, Dubna, 141980, Russia}

\affiliation[c]{Instituto de Ciencias Nucleares, Universidad Nacional Autónoma de México, Apartado Postal 70-543, Coyoacán, 04510, CdMx, Mexico}

\affiliation[d]{Facultad de Ciencias Físico-Matemáticas, Universidad Autónoma de Sinaloa, Avenida de las Américas y Boulevard Universitarios, Ciudad Universitaria, Culiacán, 80000, Sinaloa, Mexico}

\affiliation[e]{Departamento de Física, Universidad Autónoma Metropolitana-Iztapalapa, Avenida San Rafael Atlixco 186, 09340, CdMx, Mexico}

\affiliation[f]{Instituto de Física y Matemáticas, Universidad
Michoacana de San Nicolás de Hidalgo, Morelia, 58040, Michoacán, Mexico}

\affiliation[g]{División de Posgrado, Facultad de Ingeniería Eléctrica, Universidad Michoacana de San Nicolás de Hidalgo, Morelia, 58040, Michoacán, Mexico }

\emailAdd{herrera.maribel@outlook.com}

\abstract{We present the design of the mechanical structure of the mini Beam-Beam detector, a subsystem of the Multi-Purpose Detector, soon to enter into operation at the Nuclotron based Ion Collider fAcility of the Joint Institute for Nuclear Research.
The miniBeBe detector was designed and is currently being developed by the Mexican team of the NICA Collaboration to contribute to the level-zero trigger of the Time of Flight Detector. The  mechanical structure meets the requirements of minimizing the material budget and be free of ferromagnetic materials, without compromising its robustness. The design also allows for easy module replacement for maintenance and overall removal at the end of the first stage of the experiment, without affecting the rest of the subsystems, to leave room for the installation of the Inner Tracking System. In addition, a Finite Element Method analysis of the mechanical components under load was performed. Based on this analysis, it was determined that the design meets the space constraints within the Multi-Purpose Detector, as well as a deformation of less than 1 mm with overall stress of less than 2 MPa, such that no material used in the design is at risk of mechanical failure during operation. The heat transfer analysis of the cooling system revealed that the temperature of the cooling plate is maintained within a range of $19.00^{\circ}$C to $21.41^{\circ}$C, which is sufficient to ensure that the silicon photomultipliers operate below a temperature of 25.00$^{\circ}$C, thereby optimizing their functionality.}


\arxivnumber{2408.00556} % Only if you have one

\begin{document}
\maketitle
\flushbottom

\section{Introduction}
\label{introduction}

The properties of strongly interacting matter under the extreme conditions of high temperature and density have become one of the most important research venues at the forefront of fundamental research. Some experiments are designed to operate at the energies where the largest baryon density can be achieved. One of them is the Multi-Purpose Detector (MPD), currently being installed at the Nuclotron based Ion Collider fAcility (NICA) of the Joint Institute for Nuclear Research. The MPD at NICA (MPD-NICA) is designed to explore the baryon-rich
region of the quantum chromodynamics (QCD) phase diagram by colliding heavy nuclei in the center-of-mass energy per nucleon pair for heavy-ion collisions, denoted as $\sqrt{s_{NN},}$ in the range $\sqrt{s_{NN}} = 4 - 11$ GeV ~\cite{MPD:2022qhn,Golovatyuk:2019rkb,Kekelidze:2017ual}. The experiment is planned to enter in operation in a fixed target mode in the end of 2025. 

The miniBeBe detector\footnote{Its name is derived from the acronym \lq\lq Beam-Beam" counter and due to its small size, the prefix \lq\lq mini" was added.} intends, as one of its purposes, to contribute to the level-zero trigger signal for low to high multiplicity events for the Time of Flight (TOF) system through an array of Plastic Scintillators (PS) and Silicon Photo-Multipliers (SiPMs), which have proven to be useful for particle detection in nuclear, high-energy and other branches of physics~\cite{SIMON201985}. Similar trigger systems have been proposed for the MPD-NICA experiment~\cite{YUREVICH2018294,YUREVICH2020}. The detector has been designed and is currently being developed by the Mexican team of the NICA Collaboration (MexNICA).  
The first conceptual proposal of the miniBeBe version 1.0 was reported in~\cite{Kado:2020evi}.

\begin{figure}[!ht]
  \centering
 \subfloat{ \includegraphics[width=0.5\linewidth]{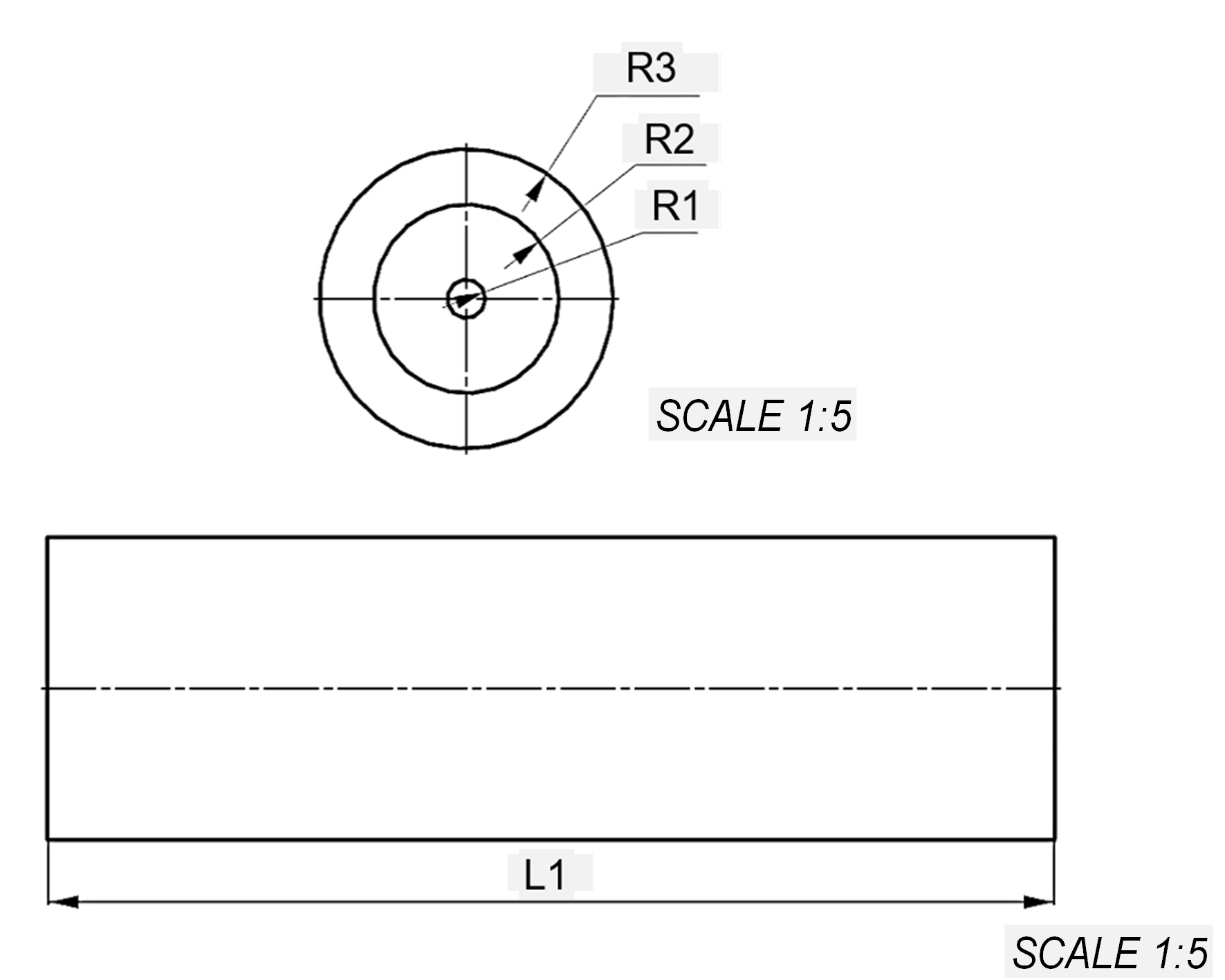}
 % \rule{6.4cm}{3.6cm} %% <- line 15
 }
 \subfloat{
\begin{tabular}{c}
    \noalign{\hrule height 0.5pt}
          \hline
          R1, beam pipe radius \\
          R2, miniBeBe detector radius \\
         R3, ITS detector radius \\\\
         \noalign{\hrule height 1pt}
         \hline 
       R1 = 32 $\pm{0.1}$ mm \\
          R2 = 156 $\pm{0.1}$ mm \\
         R3 = 247.5 $\pm{0.1}$ mm  \\
          L1 = 1160 $\pm{0.1}$ mm \\\\
        \noalign{\hrule height 1pt}
         \hline
    \end{tabular}
 }
 \caption{Geometrical constraints and tolerances for the design of the mechanics of the miniBeBe detector.}
 \label{Figure_1}
\end{figure}

\begin{figure}[!ht]
\centering
\includegraphics[width=0.9\textwidth, angle=0]{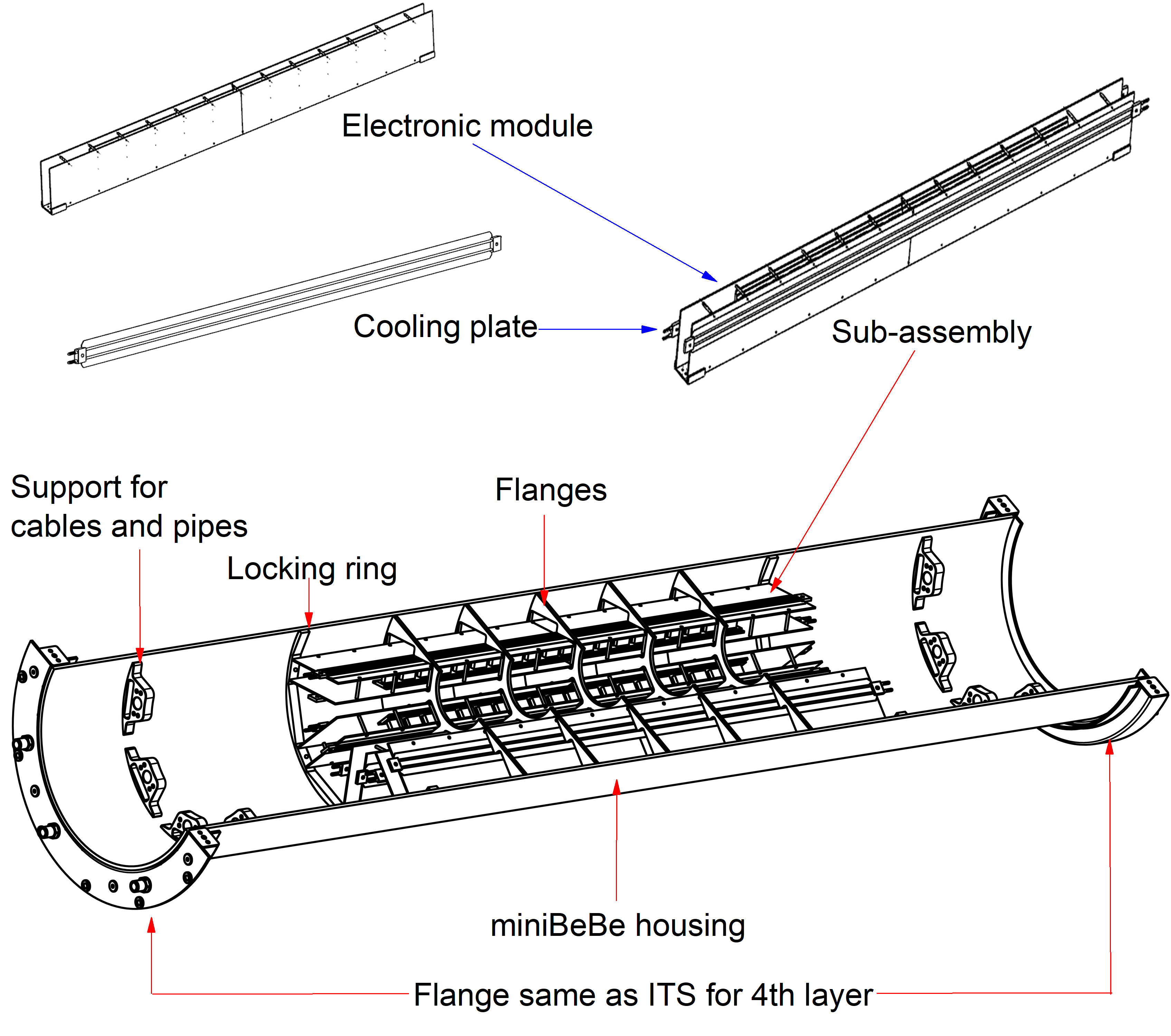}
\caption{Mechanical design concept version 2.0 of the miniBeBe, with flanges similar to the 4th layer of the ITS at the ends.}
\label{Figure_2}
\end{figure}

To realize the miniBeBe mechanical design concept, it was necessary to consider the geometric limitations of the Inner Tracking System (ITS)~\cite{Murin:2021szk}, since both detectors occupy the same space at different experimental stages. It is anticipated that the miniBeBe detector will be operational during the initial phase of the MPD experiment. The geometrical constraints and tolerances for the design are illustrated in Fig. \ref{Figure_1}, which depicts the designated detector space. The space consists of concentric cylinders, which encompass all components of the design. Figure \ref{Figure_2} shows a representation of the mechanical design concept of the miniBeBe detector.

An essential component of the MPD is a solenoid magnet with a superconducting NbTi coil and a steel flux return yoke. The magnet is designed to provide a highly homogeneous magnetic field of 0.5 T in a 4596 mm diameter aperture to ensure a transverse momentum resolution in the range of 0.1-3 GeV/c at NICA~\cite{Bunzarov:2014rfa} . Therefore, it is paramount to avoid the use of ferromagnetic materials in the construction of the detector, as they would be attracted by the magnet and affect both the structure and the results obtained by the electronic components. 

The first mechanical concept of the miniBeBe detector considered placing the electronic strips perpendicular to the radial direction. Four SiPMs were placed on one of the wider surfaces of the PS~\cite{Kado:2020evi}. A system of rails and flanges was designed to support and secure the electronic strips in order to be assembled within the ITS detector housing.

To decrease the SiPMs direct exposure to radiation, it was necessary to implement a new mechanical concept to consider that the SiPMs are now to be placed on the two narrower sides of the PS while their wider sides are still perpendicular to the radial direction. In the new arrangement, two electronic strips are in contact with the SiPMs of one PS, one at each of the narrower sides. Since it is not possible anymore to continue using the rail geometry of the first concept, where all the SiPMs were in contact with a singe electronic strip~\cite{Kado:2020evi}, a new version of the mechanical concept needed to be designed. 

A recent study~\cite{Stoykov:2021ylp} indicates that when a significant fraction of the emitted scintillation light is collected by
the photosensors, an array of PS-SiPMs, with an appropriate read-out circuit, can achieve time resolutions around $10 - 20$ ps for the detection of charged particles with energies close to 1 MeV. This kind of time resolution, and even one a bit larger, would allow us to use the miniBeBe detector as a trigger.

A proper structural design is critical to subsequently manufacture each instrument element. A good design ensures that the instrument elements are held in position, avoiding risk areas for fracture, component displacement, or unwanted vibrations that could lead to failure. The design should also determine the optimal operating range under the influence of external conditions, in particular temperature.

The requirements for an adequate design of instruments and/or detection systems have been extensively documented~\cite{MARZULLO2024114477,MEDRANO2021112651,JENIS2023538}. An example is provided by the design of the Future Circular Collider~\cite{Boscolo:2023hwo}, which serves as an inspiration to organize this work and present the mechanical design of the miniBeBe detector.

The miniBeBe will be placed inside the bore of the MPD Time Projection Chamber (TPC) using the ITS Installation Container (IC) that will slide into the TPC with a 4 mm gap. Therefore, the design of the miniBeBe mechanical structure should consider placing all the detector elements on a housing inside the IC in an orderly and straight manner. In addition, the deformation of the housing needs to be less than $1$ mm both during insertion and removal, including the loads on the detector from the internal components described in Section 3. Therefore, performing a study of high shear zones and deformation turns out to be a high priority. %to make sure that the design meets the objectives which it was conceived for.

\section{Requirements for the conceptual mechanical design}

The mechanical design concept is restricted by the geometrical constraints, illustrated in Fig. \ref{Figure_1} and by the requirement of using suitable materials for the fabrication of the detector components. The mechanical properties were obtained from the technical data sheets provided by the material suppliers and from the ANSYS 2022 library, see Table~\ref{Table_1, Materials}. The simulation packages used were Static Structural~\cite{ansys2022static} and Fluent~\cite{ansys2022fluent}. %The weight of the elements that make up the detector is calculated using the density of the materials from the technical data sheets and the volume of each part from its 3D mechanical design, ANSYS Static Structural and Fluent modules are used for the simulations.
\begin{table}[!ht] 
    \centering\caption{Suggested materials for detector construction. Properties obtained from commercial material data sheets, Young modulus (E) and density (v).} 
    \begin{tabular}{c|c|c|c|c}
    \noalign{\hrule height 1pt}
          \hline
         Material & Type & E (GPa) & v (kg / m$^{3}$) & Ref. \\
        \noalign{\hrule height 1pt}
         \hline 
        Carbon fibre& M55-J & 290 & 1910 & \cite{Carbonfiber} \\
         Polyethylene & UHMW PE & 1.1 & 950 & \cite{Polyethylene} \\
         Polystyrene & HIPS & 3.1 & 1050 & \cite{Polystyrene} \\
         Polyester film & Mylar & 4.9 & 1390 & \cite{Mylar}\\
        Aluminum & Alloy 5052 & 70 & 2680 & \cite{Aluminium} \\
         \noalign{\hrule height 1pt}
         \hline
    \end{tabular} 
   \label{Table_1, Materials}
\end{table}

The detector has been designed having in mind the use of commercially available materials in order to facilitate the search for suppliers capable of manufacturing the final components using established technological processes. The parts that make up the detector, in the order in which they are assembled, are: housing, flanges, locking ring, sub-assembly electronic module-cooling plate system and cable support. Assembly details are described in Section 4.

\section{Modeling of the mechanical concept version 2.0 of the miniBeBe detector}

\subsection{Housing}

\begin{figure} [!ht]
    \centering
    \includegraphics[width=0.65\textwidth, angle=0]{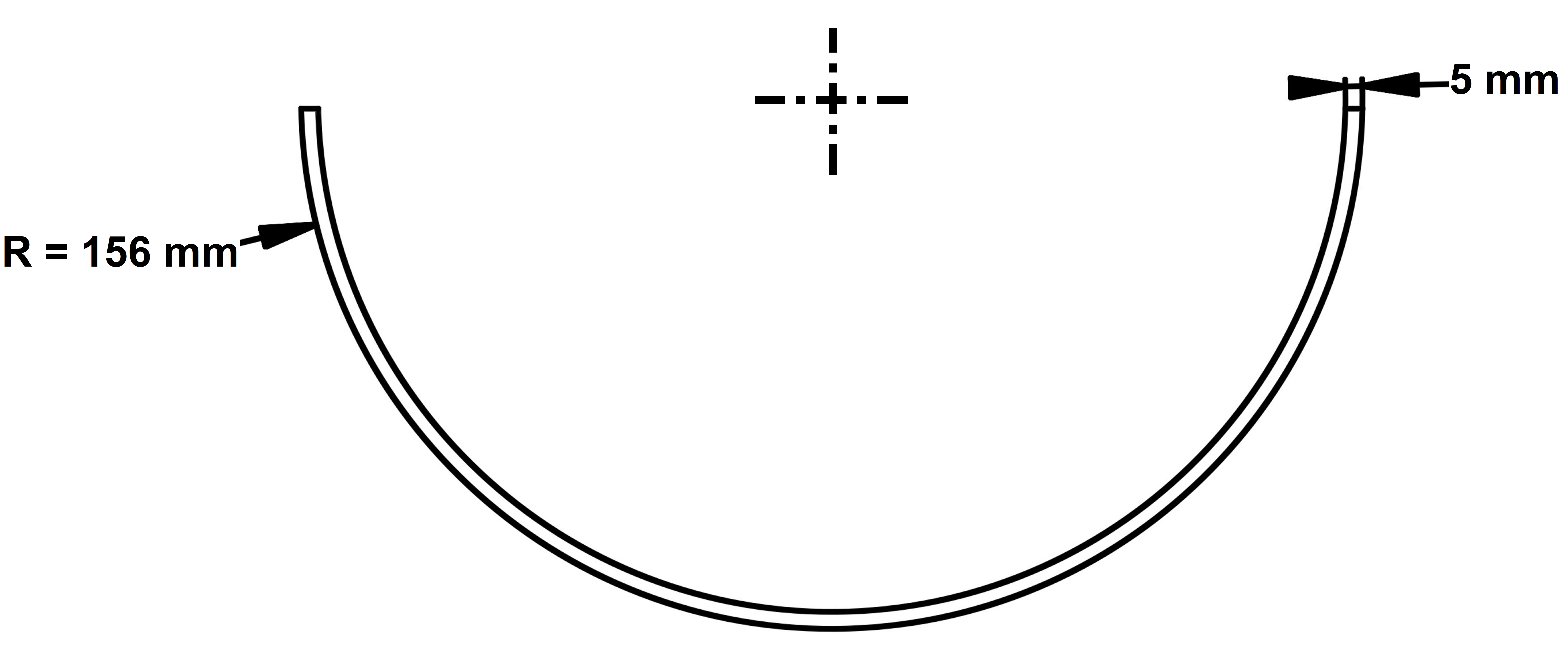}
    \caption{Front view of the miniBeBe detector housing, manufactured with M55-J as explained in the text.}
    \label{Figure_3}
\end{figure}

The housing provided by the ITS for placing the miniBeBe (see Fig.~\ref{Figure_3}) consists of a cylinder with an outer diameter of 312 mm with a thickness of 5 mm and a length of 1611 mm. The manufacturing material is M55-J~\cite{Carbonfiber}.

\begin{figure}[!ht]
    \centering
    \includegraphics[width=0.65\textwidth, angle=0]{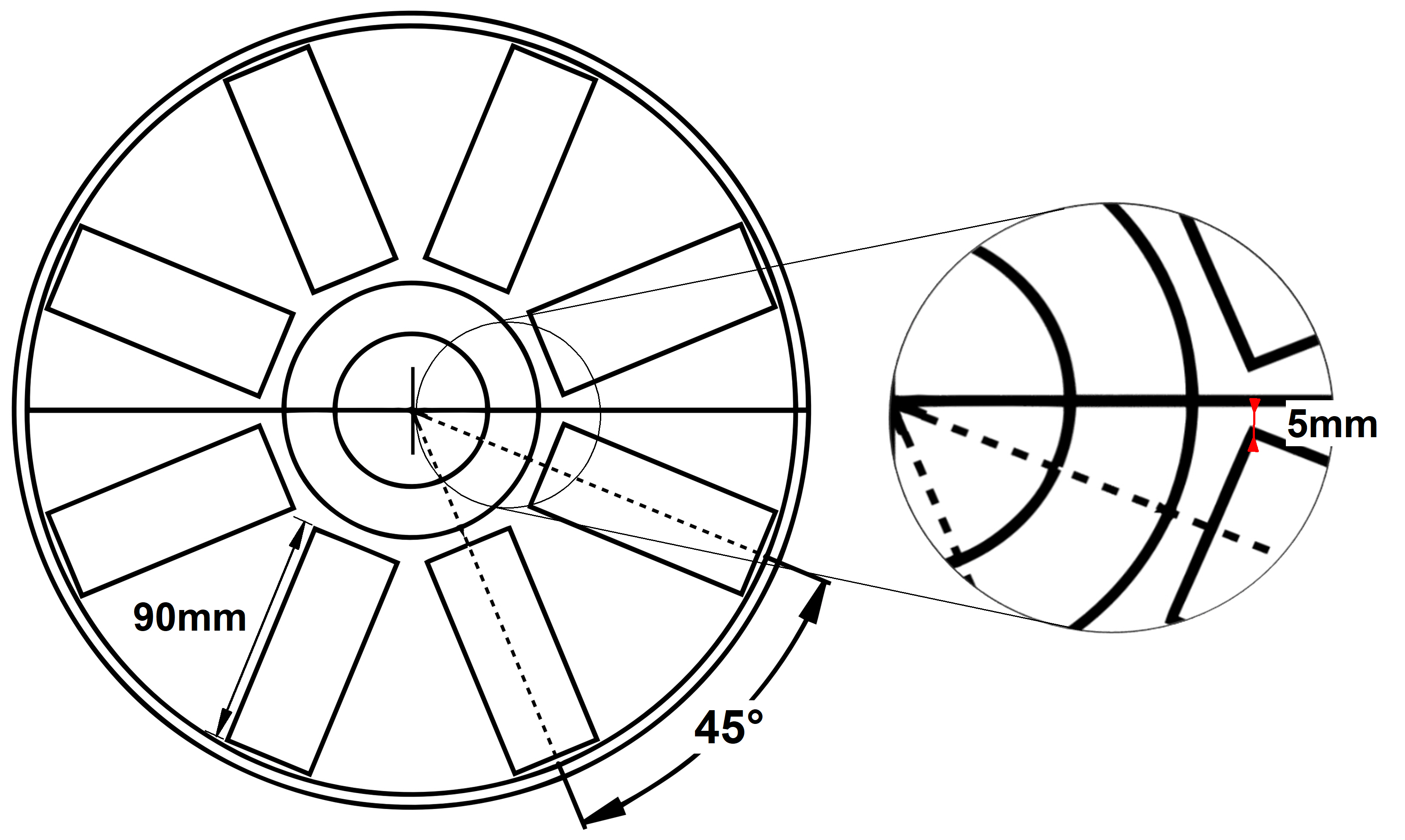}
    \caption{Transverse view of the spatial distribution of 8 modules inside the detector housing.}
    \label{Figure_4}
\end{figure}

According to these geometric parameters obtained by the electronic design group and the dimensions available to build the case, as illustrated in Fig.~\ref{Figure_1}, the maximum number of electronic strip pairs, to which the PS-SiPMs combinations are attached, is 8. This is illustrated in Fig.~\ref{Figure_4}. Hereafter, we refer to this array as the \lq\lq modules". The number of  electronic modules was also constrained by the dimensions of the Printed Circuit Board (PCB); a wide PCB reduces the space between the vertices, with the consequent reduction of the space available to place screws in a cover. In this instance, the electronic version 1.0 requires a width of 90 mm, which leaves insufficient space for screws. Therefore, the use of screws is not feasible. This is the reason why a flange system of one piece is proposed. The modules slide from one end of the mechanical structure to the clamping rings.

\subsection{Electronic module support}

\begin{figure}[!ht]
\centering
\begin{subfigure}{0.5\textwidth}
    \centering
    \includegraphics[width=1\textwidth, angle=0]{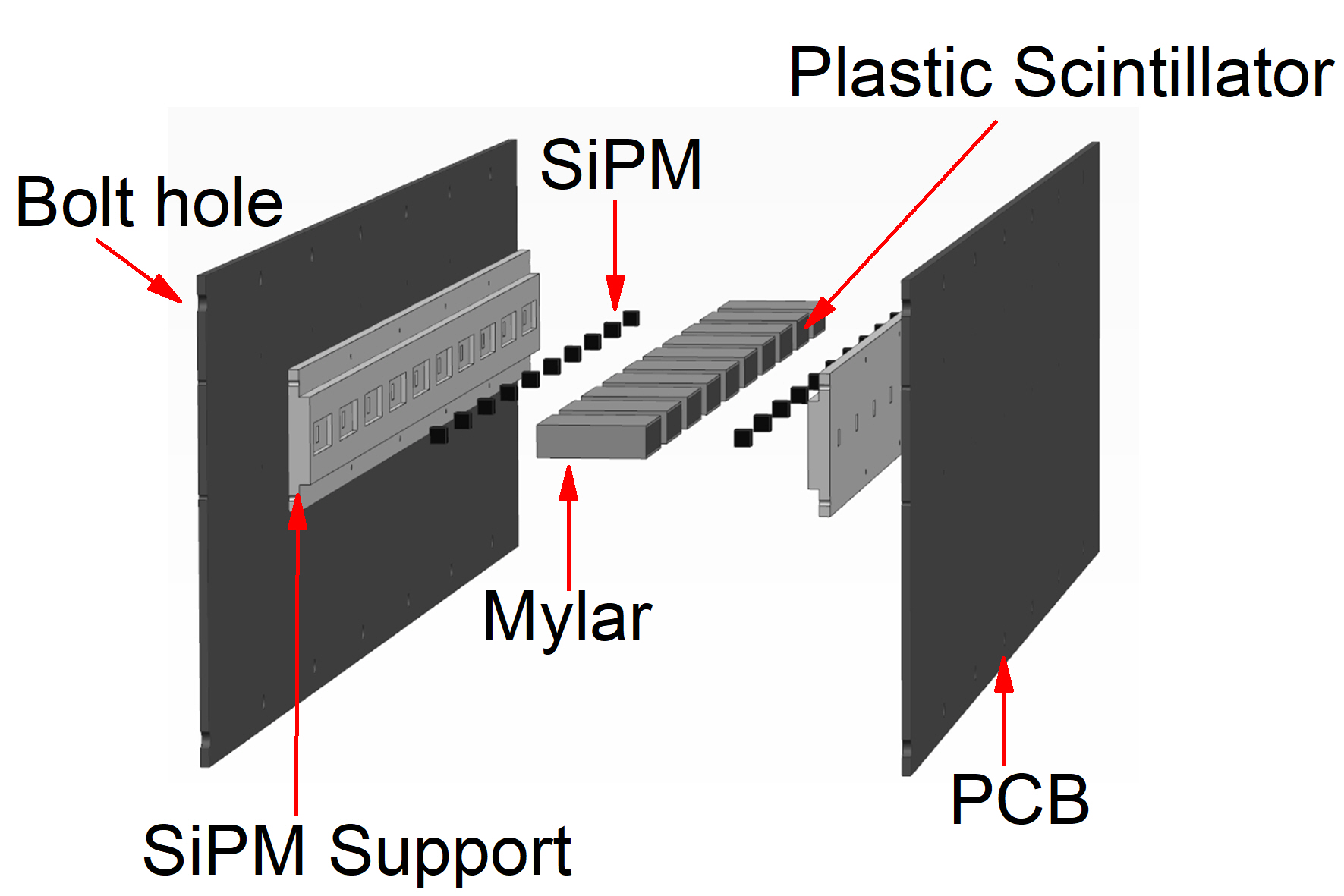}\\
    {\small (a)} 
  \end{subfigure} \\
  \begin{subfigure}{0.25\textwidth}
    \centering
    \includegraphics[width=1\textwidth, angle=0]{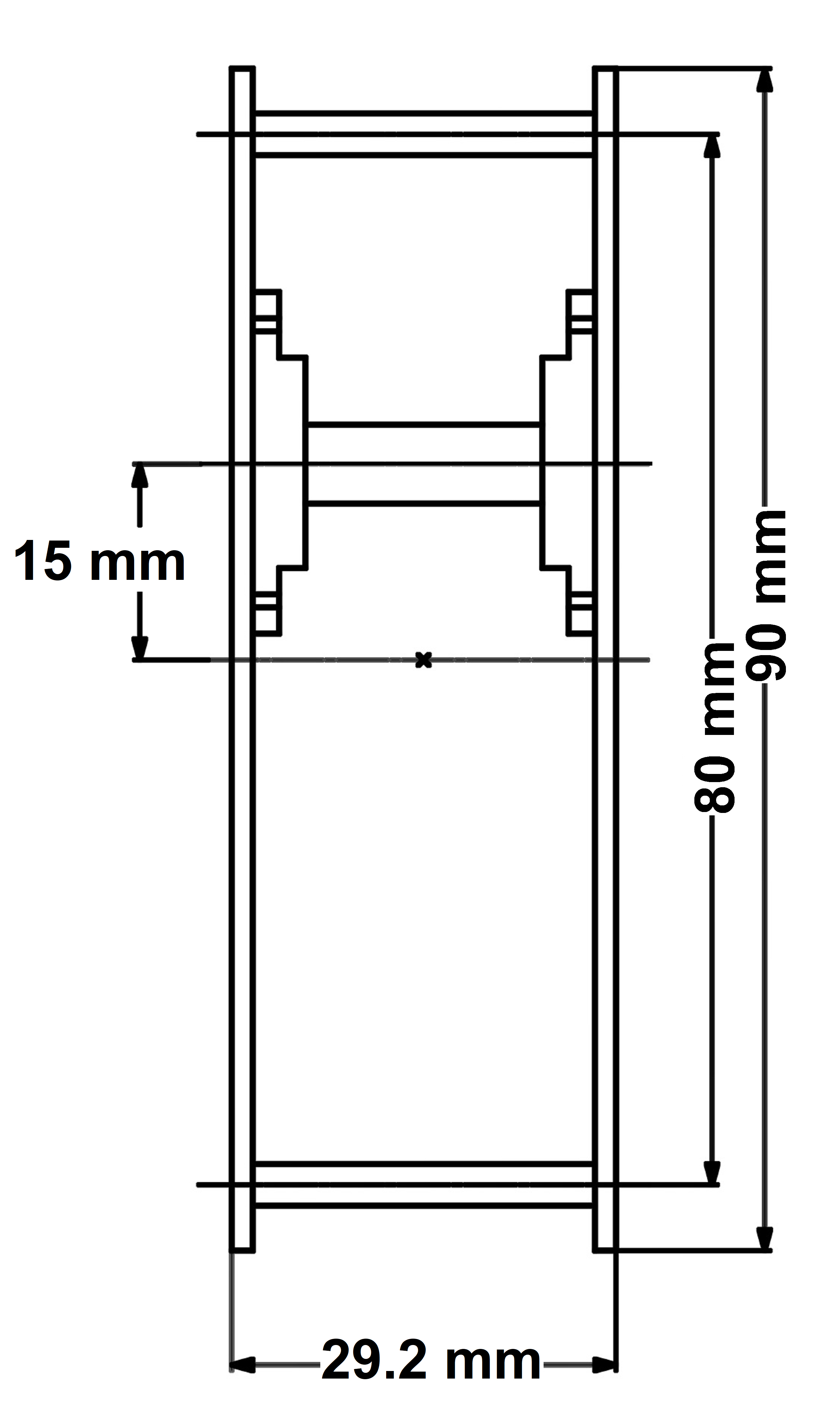}\\
    {\small (b)} 
  \end{subfigure} 
  \begin{subfigure}{0.3\textwidth}
    \centering    
    \includegraphics[width=0.5\textwidth, angle=0]{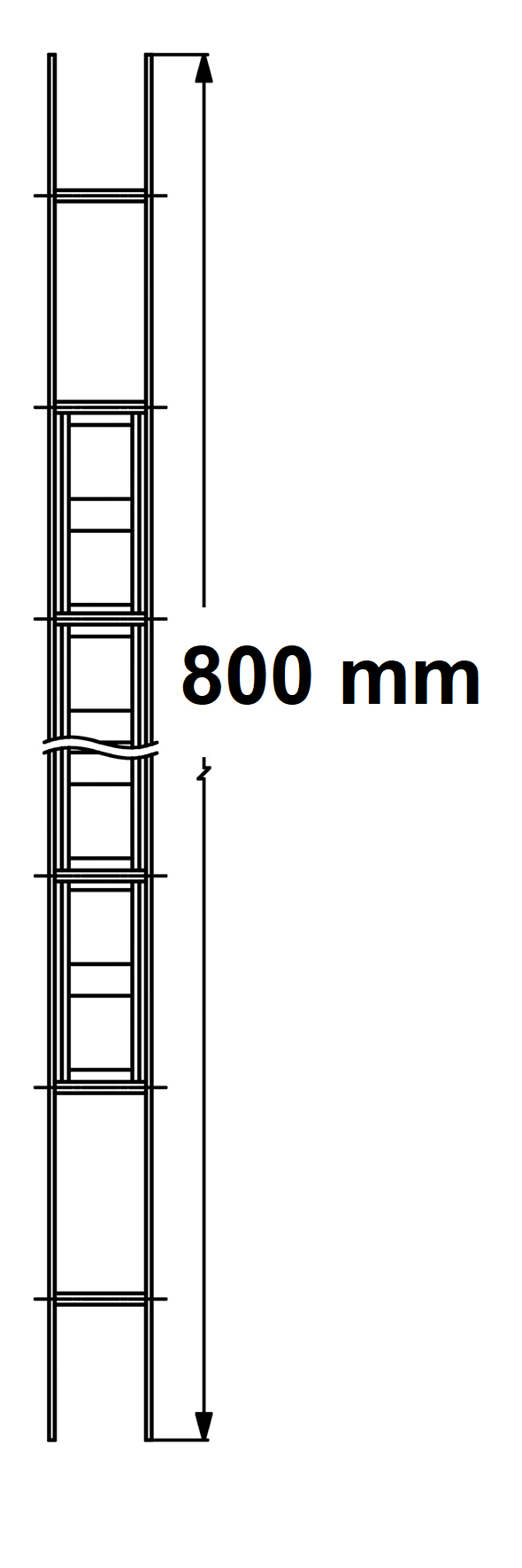}\\
    {\small (c)} 
  \end{subfigure}
  \caption{Electronic module support: (a) exploded view, (b) front view, (c) top view.}
\label{Figure_5}
\end{figure}

The electronic concept considers the need for a mechanical structure or support for each PS to be in contact with two SiPMs without needing an adhesive. An exploded view of an electronic module is provided in Fig. \ref{Figure_5}a. The cross-section of the set of electronic components is rectangular, as illustrated in Fig.~\ref{Figure_5}b, where the PS is covered with approximately 0.5 mm of Mylar and is placed at 15 mm from the center of the module. The distance between each PS edge is 10 mm. For the assembly, we consider using bolts to fix the PCBs and building a carbon fibre support to couple the SiPM-PS-SiPM set. The arrangement has a length of 800 mm, as illustrated in Fig.~\ref{Figure_5}c. A base support of 40$\times$30.5~mm$^2$  is attached to each end of the module, which is connected to a locking ring to secure the module to the detector housing. Table~\ref{Table_2} shows the number of components required to build each electronic module.

\begin{table} [!ht] 
     \centering
     \caption{Components required to build a single electronic module together with its support.}
    \begin{tabular}{c|c|c|c|c}
    \noalign{\hrule height 1pt}
          \hline
         Component - Material & Type & Volume (m$^{3}$) & Quantity & Mass (g) \\
          \noalign{\hrule height 1pt}
         \hline 
        SiPM - Polystyrene & S13360-3050PE & 2.70e-08 & 40 pcs & 1.13 \\
        Plastic scintillator - Polystyrene & EJ-232 & 2.31e-07 & 20 pcs & 4.86\\
        SiPM support - Carbon fibre  & M55-J & 5.74e-05 & 2 pcs & 438.70\\
       PCB - Polyethylene  & FR-4 & 2.37e-05 & 2 pcs & 89.89 \\
       Mylar - Polyester film & WC-Film & 5.20e-07 & 2$\times$100 cm$^2$ & 14.46\\
       Module support - Carbon fibre & M55-J & 1.51e-06 & 2 pcs & 5.77\\
        Bolts - Carbon fibre & M55-J & 2.00e-06 & 26 pcs & 99.32\\
         \noalign{\hrule height 1pt}
         \hline
        & & & Total mass & 654.12\\
        \noalign{\hrule height 1pt}
         \hline
         & & & + 25 percent safety & 817.65\\
        \noalign{\hrule height 1pt}
         \hline
    \end{tabular} 
    \label{Table_2}
\end{table}

\begin{figure}[!ht]
\centering
  \begin{subfigure}{0.5\textwidth}
    \centering
    \includegraphics[width=1\textwidth, angle=0]{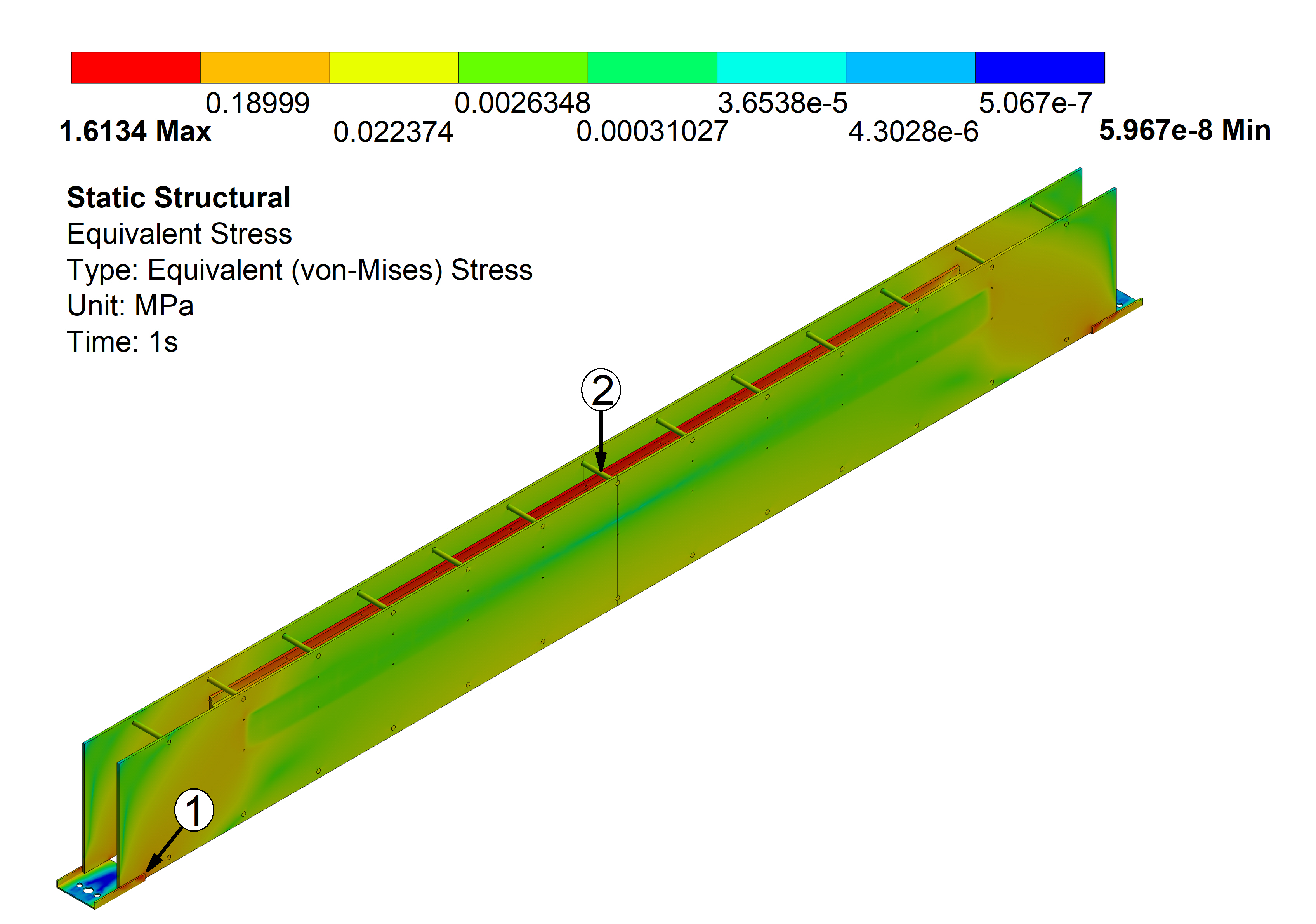}\\
    {\small (a)} 
  \end{subfigure}
    \begin{subfigure}{0.45\textwidth}
    \centering    
    \includegraphics[width= 1\textwidth, angle=0]{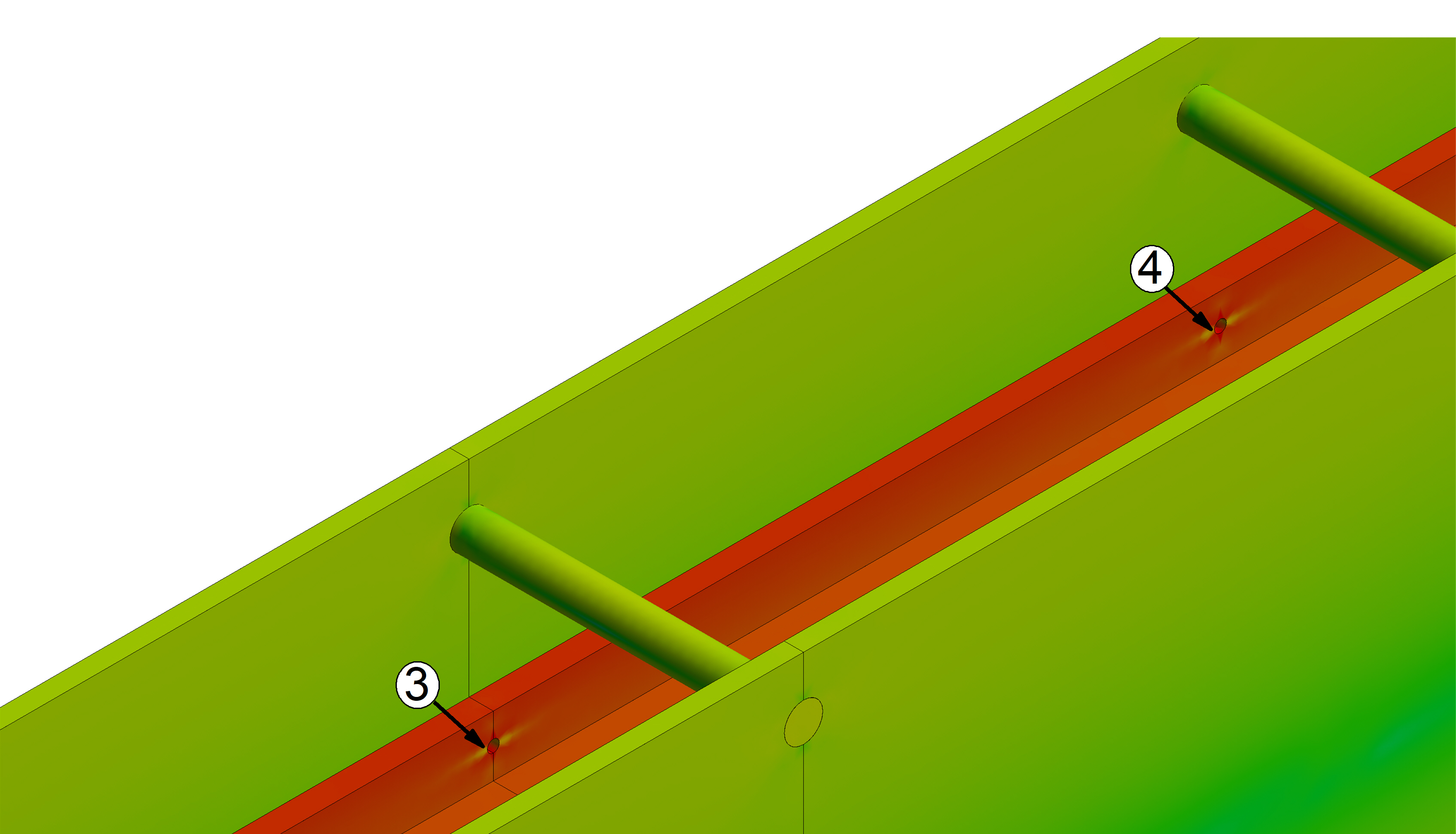}\\
    {\small (b)} 
  \end{subfigure}
  \caption{Analysis of the electronic module and of the base support: (a) isometric view and (b) close-up view.}
\label{Figure_6}
\end{figure}

For the finite element analysis of the electronic module, the support bases were considered to support the weight of each module and its components. Figure~\ref{Figure_6} shows the zones of highest failure risk. Table~\ref{Table_3} shows the stress and strain values corresponding to these zones. In addition, the shear stress between the SiPMs support and the PCB is 0.70 MPa, so there is no risk of support detachment.

\begin{table} [!ht] 
    \centering
    \caption{Values of stress and deformation of the electronic module. The areas of maximum stress are indicated in Fig.~\ref{Figure_6}.}
    \begin{tabular}{c|c|c}
    \noalign{\hrule height 1pt}
        \hline
         Zone & Von Mises (MPa) & Deformation (mm) \\
          \noalign{\hrule height 1pt}
         \hline 
         1 & 0.258 & 1.837e-04\\
         2 & 0.513 & 1.032e-02\\
         3 & 0.034 & 1.031e-02\\
         4 & 0.029 & 1.007e-02\\
         \noalign{\hrule height 1pt}
         \hline
    \end{tabular} 
    \label{Table_3}
\end{table}

\subsection{Cooling plate}

\begin{figure}[!ht]
\centering
\begin{subfigure}{0.6\textwidth}
    \centering
    \includegraphics[width=1\textwidth]{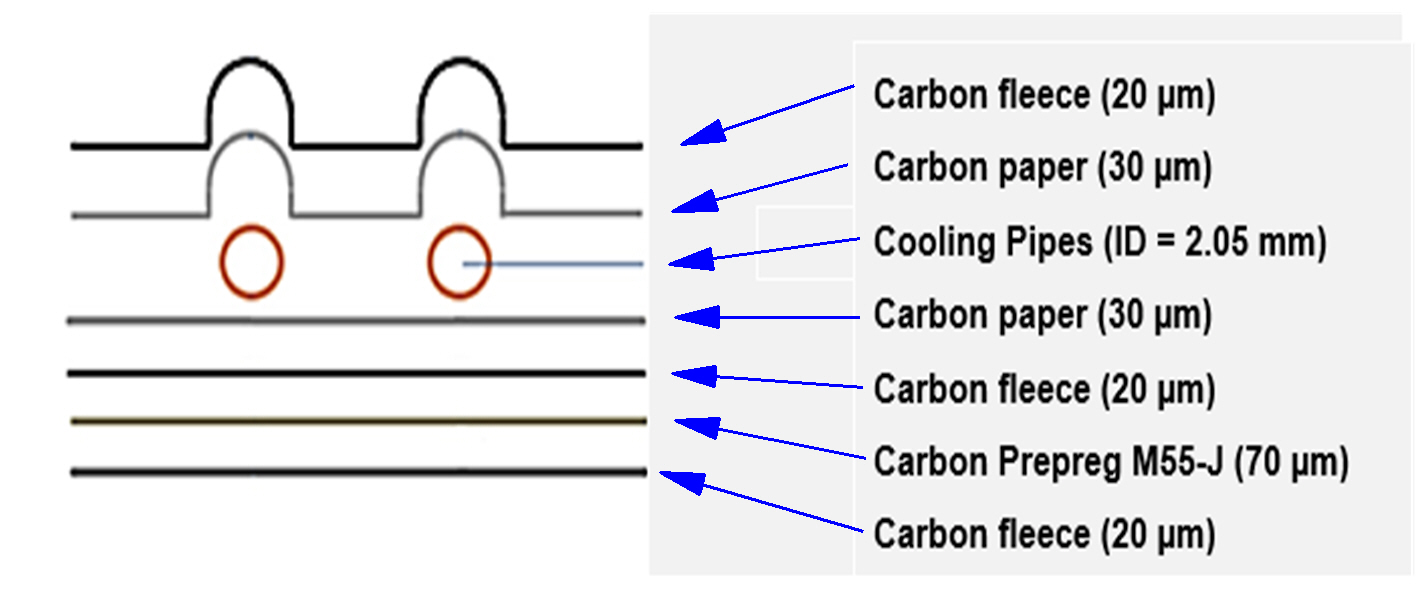}\\
    {\small (a)} 
  \end{subfigure}
  \begin{subfigure}{0.7\textwidth}
    \centering
    \includegraphics[width=1\textwidth]{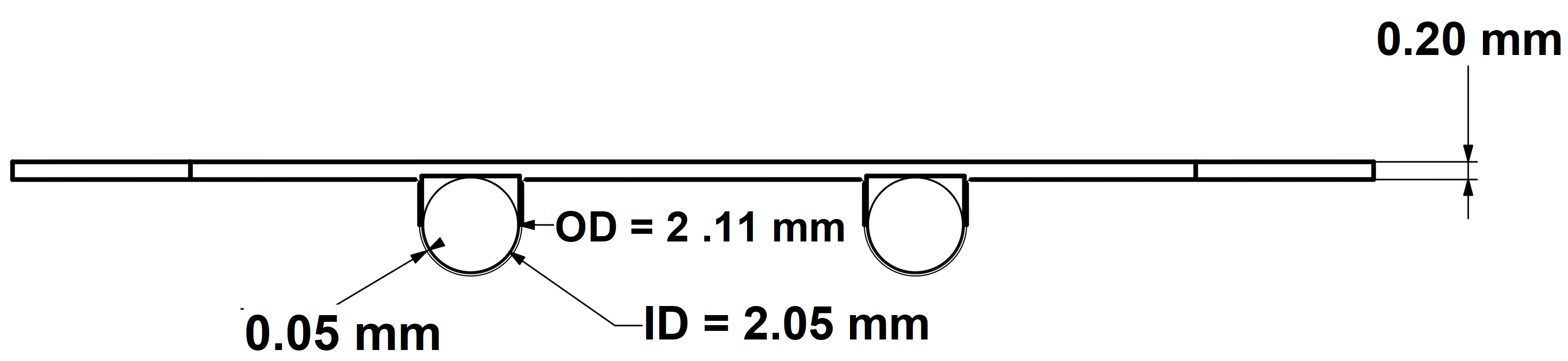}\\
    {\small (b)} 
  \end{subfigure}
  \begin{subfigure}{0.7\textwidth}
    \centering
    \includegraphics[width=1\textwidth]{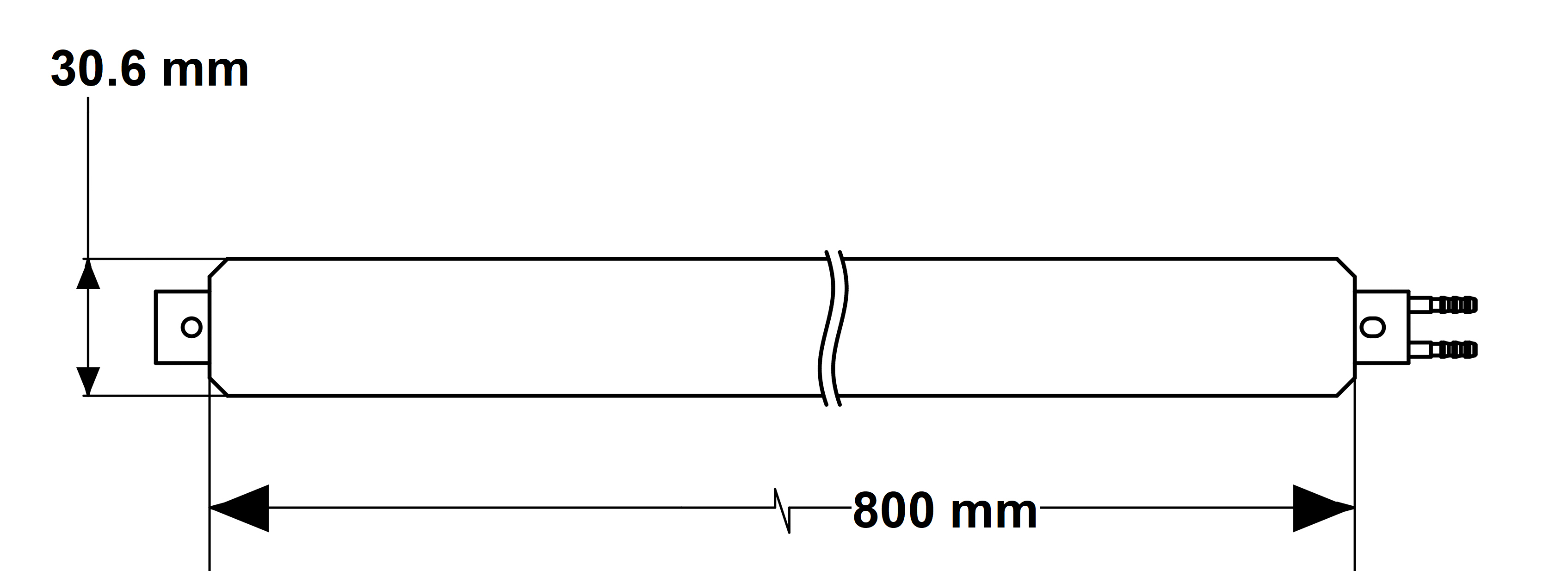}\\
    {\small (c)} 
  \end{subfigure}
  \caption{Cooling plate: (a) Composite materials for the cooling plate, (b) Front view and (b) Top view.}
  \label{Figure_7}
\end{figure}

The cooling plates to be used for removing the heat load from the modules will be provided by the ITS using the same technological process described in the patent RU2806879 \cite{Tuyana}. The plate consists of a composite of carbon fibre papers as shown in Fig.~\ref{Figure_7}a and Eccobond~45 adhesive resin, adjusted to the miniBeBe length requirements. This consists of a cold plate with two polyamide tubes embedded inside for water flow with an outer diameter of 2.11 mm and an inner diameter of 2.05 mm, the total thickness for the carbon fibre is $\sim$ 0.20 mm as illustrated in Fig.~\ref{Figure_7}b. In the specific case of the miniBeBe detector, the plate is 800 mm long and 30.6 mm wide, as illustrated in Fig.~\ref{Figure_7}c. Attachment of the cooling plate to the module is achieved through the use of Eccobond 45, with a thickness of 100 $\mu$m, this has a negligible impact on the outcome. This adhesive is employed in the construction of other detectors, including ALICE-ITS \cite{abelev2014technical}, which is utilized in high-energy environments. 

According to the experimental results obtained by the group responsible for the development of the ITS  detector~\cite{sheremetev2023mpd}, the cooling plate is able to maintain a thermal load of 40 mW/cm$^2$ in a range from $20.58^{\circ}$C to $21.76^{\circ}$C at an inlet water temperature of $18.00^{\circ}$C and a flow rate of 0.5~l/hr. A simulation of the physical experiment was carried out to determine the feasibility of using the same cooling plate for the miniBeBe detector.

\begin{figure} [!ht]
    \centering
    \includegraphics[width=0.7\textwidth, angle=0]{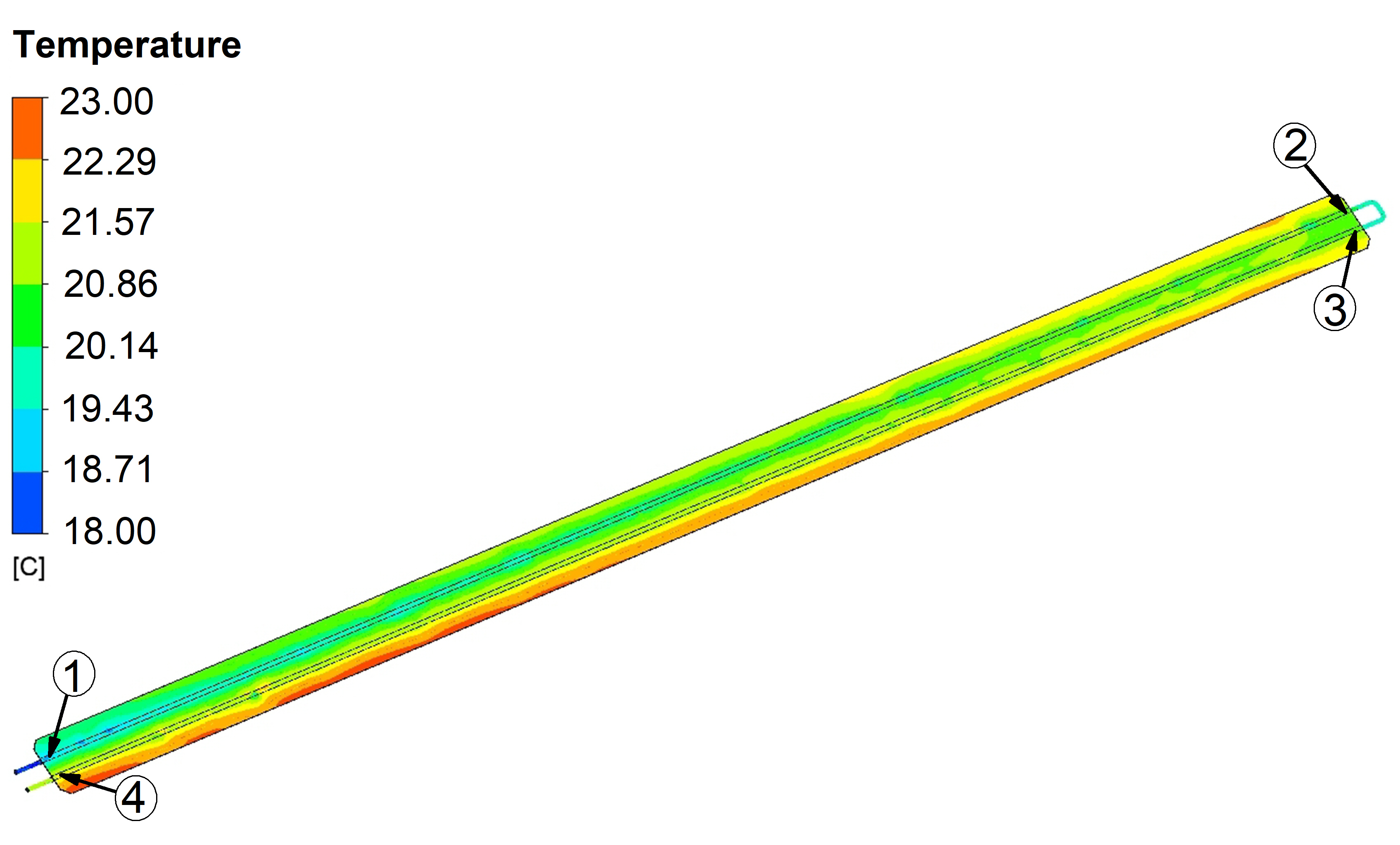}
    \caption{Cooling plate heat transfer. The plate is manufactured with a composite of carbon fibre papers Eccobond 45 adhesive resin, and the tubes of polyamide.}
    \label{Figure_8}
\end{figure}

In order to facilitate the simulation of the plate, the mechanical properties were simplified by assuming that all components are manufactured with  M55-J, given that carbon fibre constitutes the majority of the construction. According to the results of the Fluent simulation by the finite volume method, the water temperature at the outlet of the cooling plate is $21.41^{\circ}$C. The results of the analysis of the temperature distribution on the cooling plate for the miniBeBe detector is shown in Fig.~\ref{Figure_8}. Table \ref{Table_4}  displays the temperature on the cooling plate: inlet, inlet return, outlet return, and outlet  zone. The SiPMs can operate in a temperature range from $-20^{\circ}$C to $80^{\circ}$C~\cite{SiPM}. The proposed liquid cooling system is effective in stabilizing the temperature of the SiPMs in contact with the cold plate, with values below $22.66^{\circ}$C. This allows the SiPMs to operate below a temperature of $25^{\circ}$C.

\begin{table} [!ht] 
     \centering
     \caption{Temperature distribution over the cooling plate.}   
    \begin{tabular}{c|c|c}
    \noalign{\hrule height 1pt}
        \hline
         Zone & On cooling plate &Temperature ($^{\circ}$C) \\
          \noalign{\hrule height 1pt}
         \hline 
         1 & inlet & 19.00 \\
         2 & inlet-return& 20.75 \\
         3 & outlet-return& 20.72 \\
         4 & outlet & 21.41 \\
         \noalign{\hrule height 1pt}
         \hline
    \end{tabular} 
    \label{Table_4}
\end{table}

On the other hand, on the back of the plate there are high temperature zones from $30^{\circ}$C to $44^{\circ}$C. This is because there is less material on the back of the cooling plate to dissipate heat, see Fig. \ref{Figure_7}b. In addition, there is no more nearby material to dissipate heat.

As indicated in the data sheet, the melting point of the polyamide tubes is $255.00^{\circ}$C~\cite{Polyamide}, while the melting point of the resin forming part of the carbon fibre composite is $200.00^{\circ}$C~\cite{Eccobond45}. Therefore, there is no risk that the plate material will be destabilized. There is a possibility of using the ITS gas cooling system. By adding this system, the temperature is expected to be reduced by approximately $2^{\circ}$C more compared to only using a water cooling system, as was reported in Ref. \cite{TDR-ITS}.

\subsection{Flanges}

\begin{figure} [!ht]
    \centering
    \includegraphics[width=0.6\textwidth, angle=0]{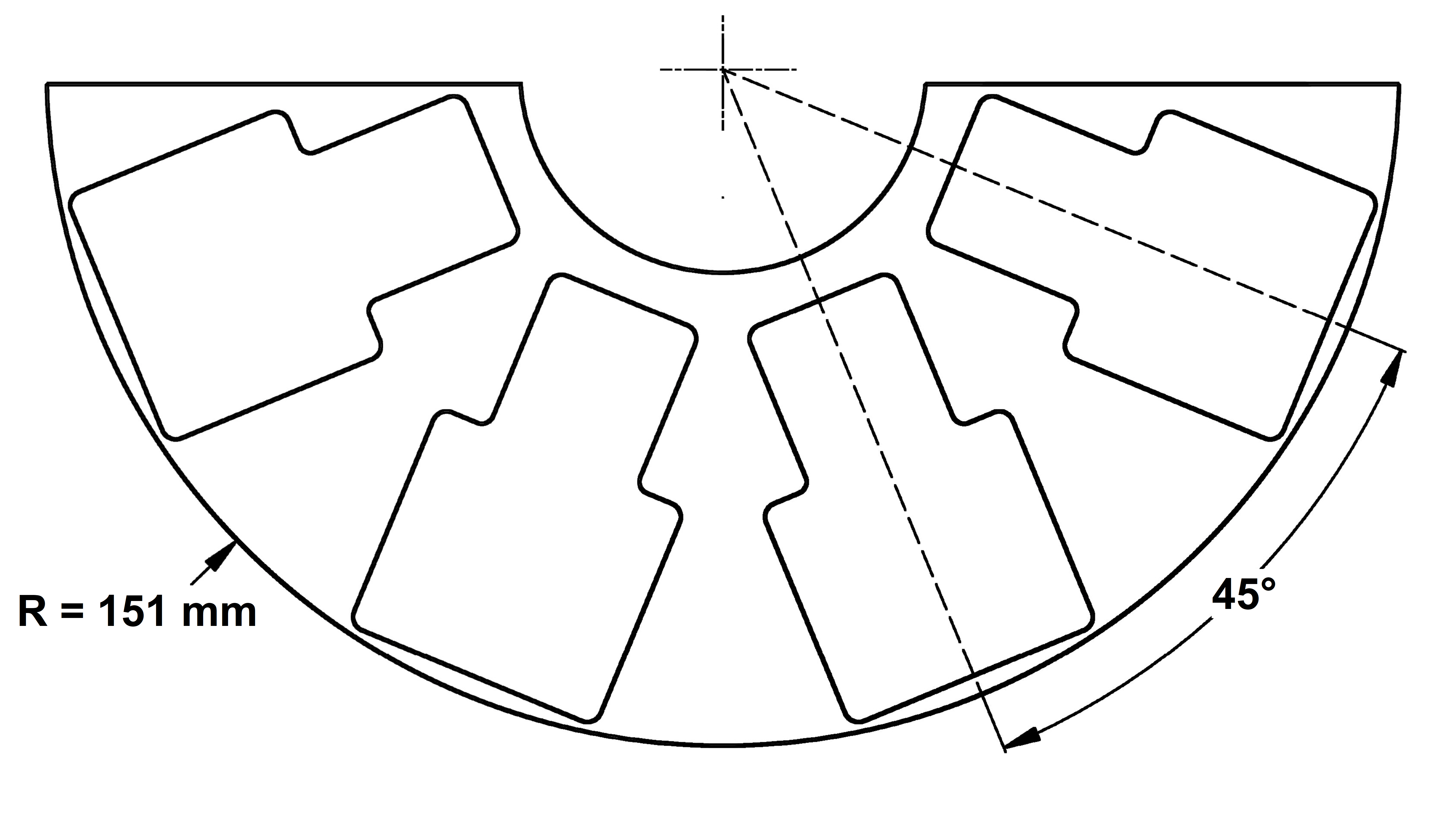}
    \caption{Front view of a flange. The manufacturing material is  aluminum alloy 5052, as explained in the text.}
    \label{Figure_9}
\end{figure}

This design, version 2.0, eliminates the use of rails of version 1.0 to use instead flanges to simplify the manufacturing process and reduce costs. The slot through which the modules move also allows for the passage of the connectors located at the ends of the modules. The slot is designed for a $50\times10$ mm$^2$ connector. The manufacturing material of the flange is aluminum alloy 5052~\cite{Aluminium}. The angular separation of the slots is $45^{\circ}$ to maintain symmetry and an even load distribution, as illustrated in Fig.~\ref{Figure_9}, and the width is 5~mm. There are a total of ten flanges, five for each half detector housing.

\begin{figure} [!ht]
    \centering
    \includegraphics[width=0.6\textwidth, angle=0]{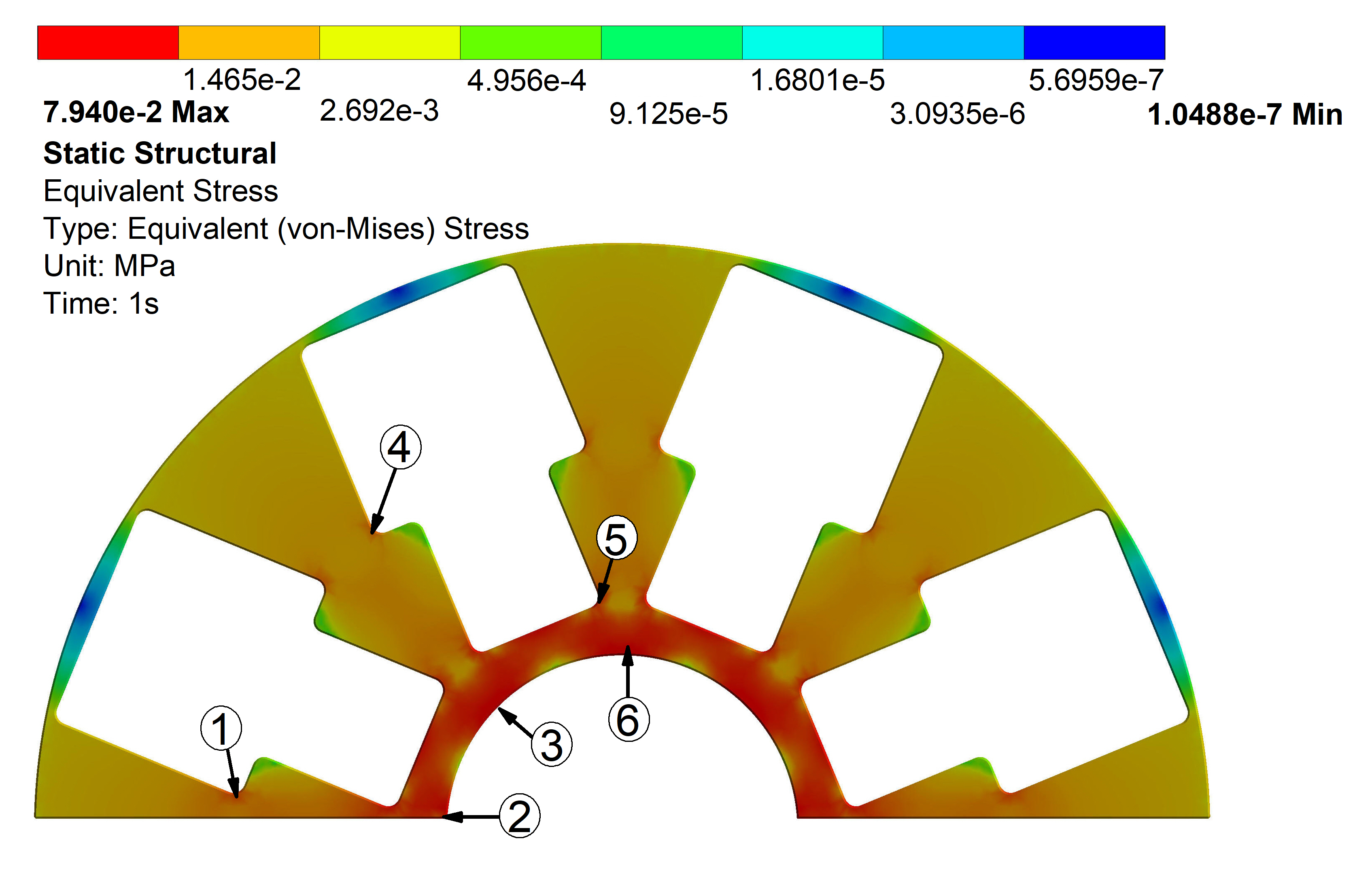}
    \caption{Flange stress analysis. }
    \label{Figure_10}
\end{figure}

The mass of the parts that make up the module assembly, illustrated in Fig.~\ref{Figure_5}, is 660 g. For the simulation, a safety factor of 25${\%}$ is considered, resulting in a module mass of 825 g. The load borne by each flange section is 1.6 N, which corresponds to the subassembly section it supports. Each subassembly crosses the window of each flange, with a total of five flanges for every half detector housing. The stress simulation was performed placing the flange on the highest risk area, which corresponds to the top of the housing. The results are illustrated in Fig.~\ref{Figure_10}.

Table~\ref{Table_5} shows the results for each of the flanges. In light of the fact that the flange design is symmetrical with respect to a vertical axis located in the central portion, it is imperative to understand that the values presented are to be interpreted as reflecting values on the opposite side of the geometry. It is observed that the stress concentration zones are located in the sections where the geometric changes happen. However, given the obtained stress values, less than 1~MPa, there is no risk of material failure and the design meets the safety requirements.

\begin{table} [!ht] 
    \centering
    \caption{Simulation results for the flange stress and deformation values in the indicated zones.}
    \begin{tabular}{c|c|c}
    \noalign{\hrule height 1pt}
        \hline
         Zone & Von Mises (MPa) & Deformation (mm) \\
          \noalign{\hrule height 1pt}
         \hline 
         1 & 0.016 & 4.027e-06\\
         2 & 0.058 & 1.308e-05\\
         3 & 0.044 & 9.163e-06\\
         4 & 0.012 & 3.284e-06\\
         5 & 0.022 & 7.742e-06\\
         6 & 0.035 & 8.979e-06\\
         \noalign{\hrule height 1pt}
         \hline
    \end{tabular} 
     \label{Table_5}
\end{table}

The total deflection of the flange in each direction is $x=$1.6500e-05 mm, $y=$1.450e-06 mm and $z=$1.890e-07 mm. Therefore, the design of the flanges is within the allowable deformation parameters shown in Fig.~\ref{Figure_1}. The assembly of the remaining detector elements is not affected by the positioning of these parts.

\subsection{Fixing supports}

The primary function of the fixing bracket is to guide and center the electronic modules, cables, and pipes feeding the cooling system, thereby avoiding load transfer to the sub-assembly and ensuring the structural integrity of the entire assembly. The mechanical design of these fixing elements is described in Section 4. 

\subsubsection{Locking ring}

\begin{figure} [!ht]
    \centering
    \includegraphics[width=0.6\textwidth, angle=0]{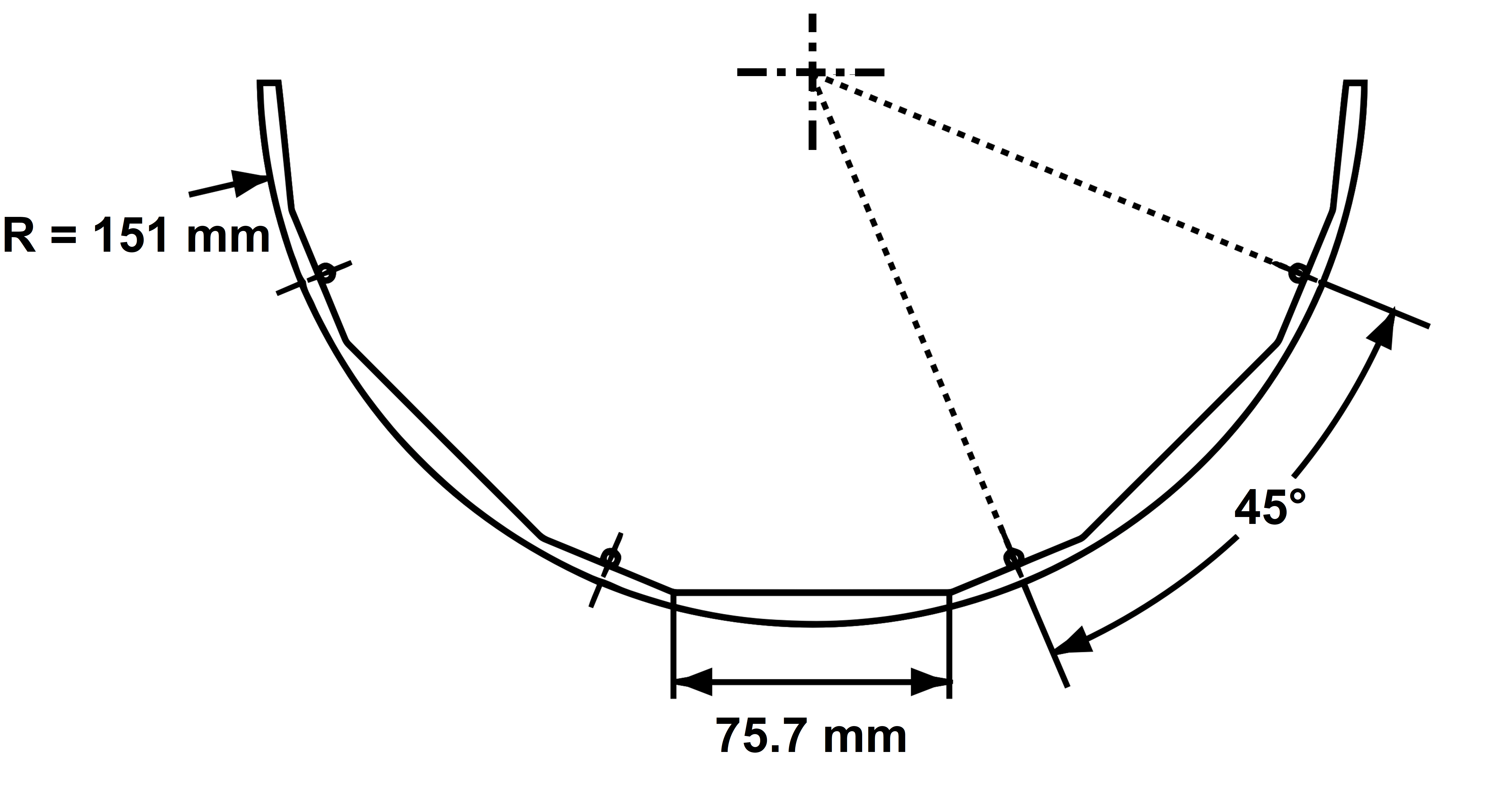}
    \caption{Front view of the locking ring. The manufacturing material is aluminum alloy 5052, as explained in the text.}
    \label{Figure_11}
\end{figure}

The design includes a locking ring between the modules and the housing, as illustrated in Fig.~\ref{Figure_11}. For this purpose, high precision ruby balls will be used to center each module in the middle of the flange slot; the radius of the ring is 151 mm and the width is 10 mm. Once in position, safety screws are used to fix it. The material of the locking ring is aluminum alloy 5052 ~\cite{Aluminium}.

\subsubsection{Support for pipes and cables}

\begin{figure} [!ht]
    \centering
    \includegraphics[width=0.6\textwidth, angle=0]{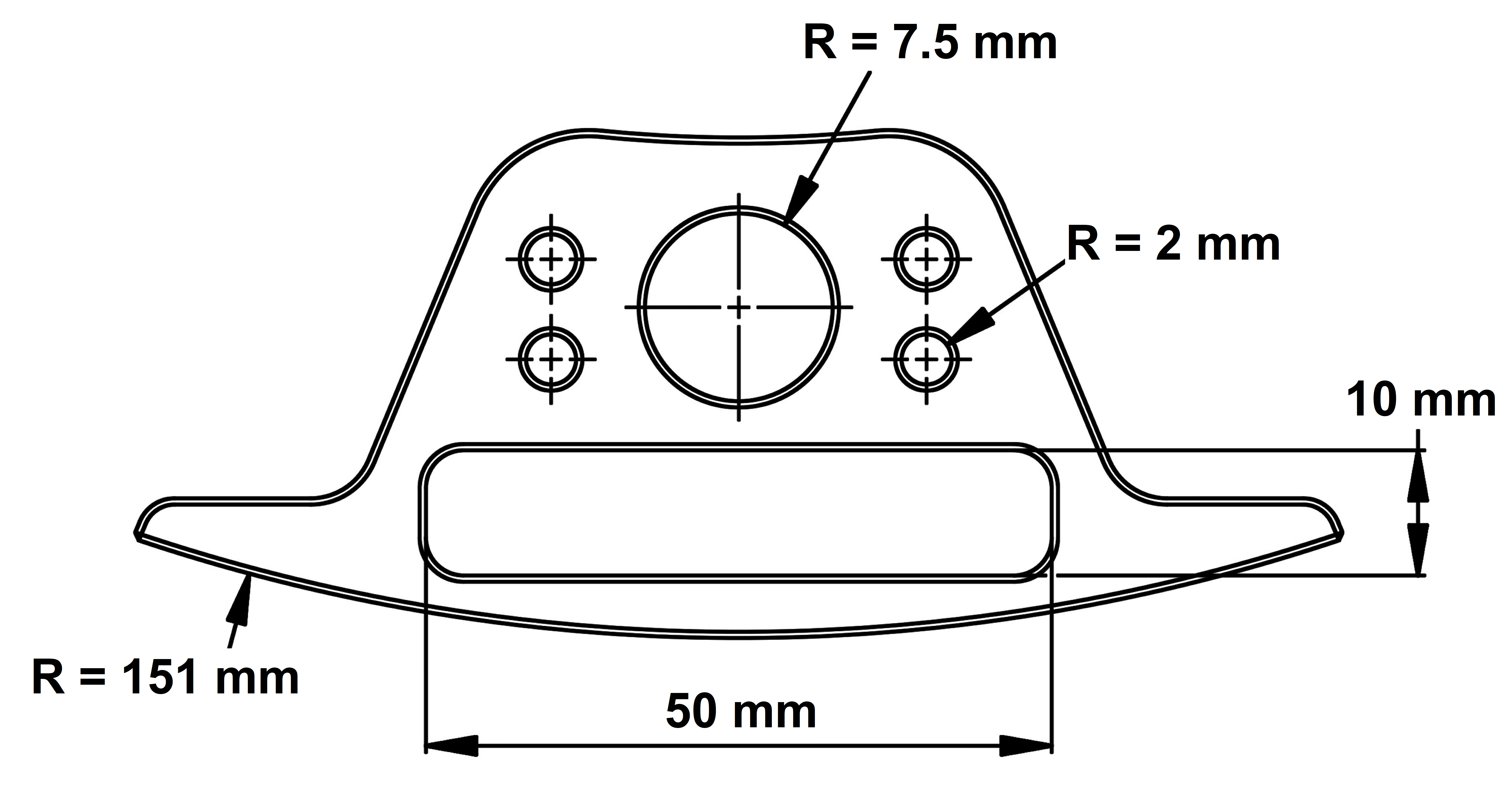}
    \caption{Front view of the support for pipes and cables. The manufacturing material is aluminum alloy 5052.}
    \label{Figure_12}
\end{figure}

To prevent cables and pipes from falling into the housing, a support has been designed, as illustrated in Fig.~\ref{Figure_12}. It consists of 4 mm holes to allow the flow of water of the cooling system (inlet and outlet), a 15 mm hole to connect the gas cooling system and a $50\times10$ mm$^2$ slot for the passage of cables to connect the modules; the radius of the support is 151 mm, the width is 12 mm and the manufacturing material is aluminum alloy 5052~\cite{Aluminium}.

\section{Assembly of the mechanical concept version 2.0 of the miniBeBe detector}

Figure~\ref{Figure_2} shows a schematic representation of the complete assembly of the miniBeBe detector version 2.0; version 1.0 was described previously~\cite{Kado:2020evi}. The main changes were the geometry and the arrangement of the electronic boards. In version 1.0, electronic strips were used to wrap the beam pipe and one SiPM was used for each scintillator plastic. In version 2.0 of the electronics, two SiPMs are needed for each PS. So, to ensure the contact among the elements, the modules were proposed precisely as shown in Fig.~\ref{Figure_5}. The figure shows four modules connected to the housing. In total, there will be eight modules inside the housing. There is a gasket between the two halves of the housing to provide a hermetic seal. If a module needs to be replaced, this can be accomplished unscrewing the cable support to slide out the entire module without affecting the rest of the assembly. The detector housing is supported by the 4th stage flanges, that correspond to the ITS.
%\begin{figure}[!ht]
%    \centering
%    \includegraphics[width=1\textwidth, angle=0]{Figure_2.jpg}
%    \caption{Mechanical design concept version 2.0 of the miniBeBe, with flanges similar to the 4th layer of the ITS at the ends.}
%    \label{Figure_2}
%\end{figure}

\begin{figure}[!ht]
    \centering
    \includegraphics[width=0.8\textwidth, angle=0]{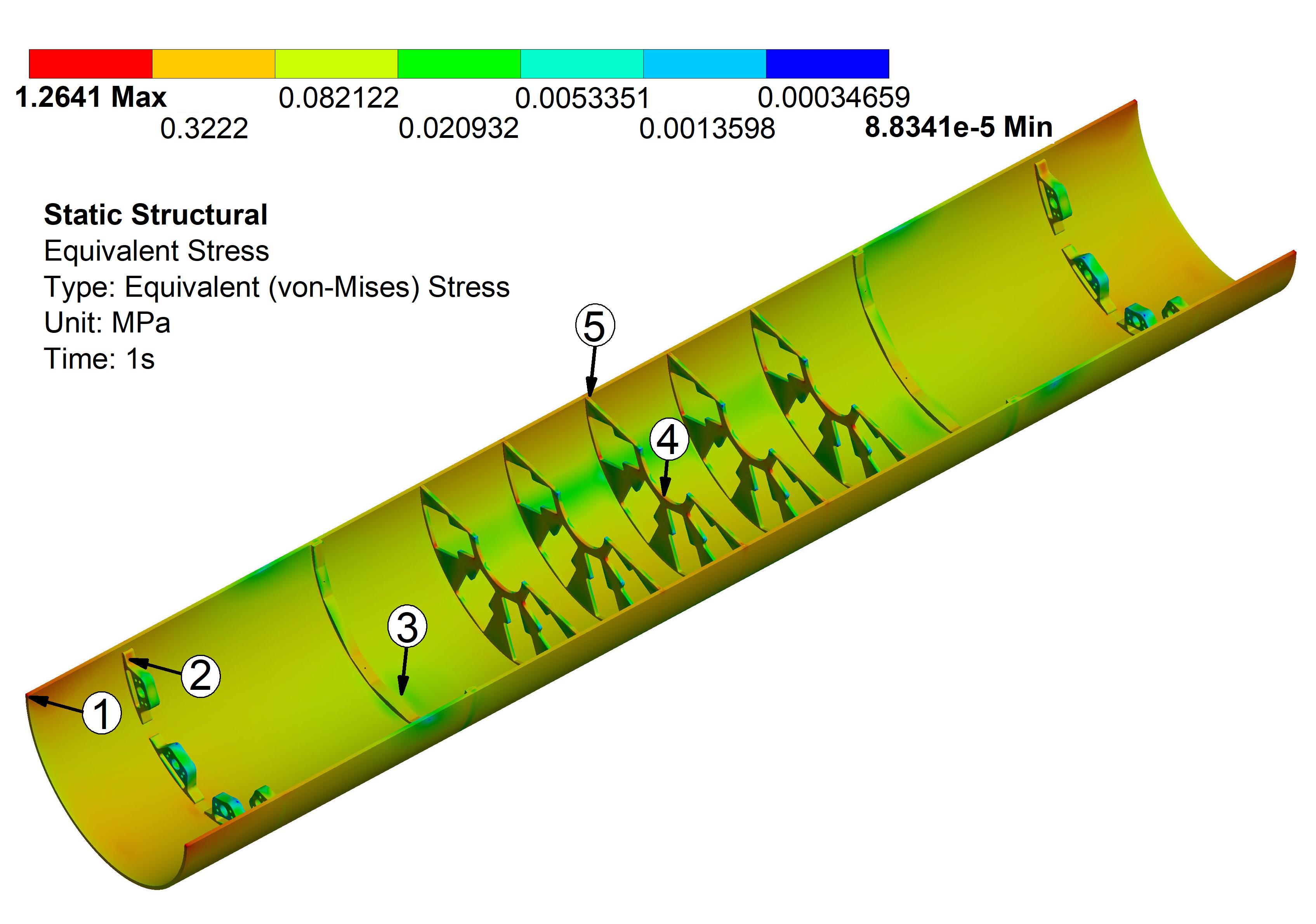}
    \caption{Mechanical design concept analysis for miniBeBe version 2.0.}
    \label{Figure_13}
\end{figure}

Finally, the behavior of the housing, with all the elements of the detector, was simulated and results are schematically shown in Fig.~\ref{Figure_13}. For the case of the modules, a load of 1.6 N was applied to the flange sections with the housing fixed at the ends. Table~\ref{Table_6} shows the value of the stress and deflection in the areas of high-stress concentration. The complete detector is cylindrical and completely symmetric with respect to its axis. The specific points analyzed are those where maximum stress is expected; additional points correspond to mirror reflections of these maximum stress points. The total deformation of the case is less than 1 mm, so the concept of the mechanical design version 2.0 is safe.

\begin{table}[!ht]
    \centering
    \caption{Stress and deformation values of the detector structure after assembly, showing the areas of maximum stress.}
    \begin{tabular}{c|c|c}
    \noalign{\hrule height 1pt}
        \hline
         Zone & Von Mises (MPa) & Deformation (mm) \\
          \noalign{\hrule height 1pt}
         \hline 
         1 & 0.459 & 4.856e-04\\
         2 & 0.165 & 7.068e-03\\
         3 & 0.175 & 8.701e-03\\
         4 & 0.223 & 5.085e-02\\
         5 & 0.327 & 5.095e-02\\
         \noalign{\hrule height 1pt}
         \hline
    \end{tabular} 
    \label{Table_6}
\end{table}

\subsection{Mass calculation and improvements to be considered}

The preceding finite element analyses demonstrate that the design meets the required deformation criteria and that the stresses within the components remain below the elastic limit of the material employed. This allows for the incorporation of further enhancements into the miniBeBe detector version 2.0 concept design report.

At this phase of the design process, it is evident that the incorporation of mass reduction strategies is a crucial aspect that can enhance the overall performance. The utilization of carbon fibre as a manufacturing material presents a viable solution to this end. Table \ref{Table_7} presents a comparison of the mass of the original aluminum components (standard design) and manufactured using carbon fibre (optimised design).The mass of the constituent elements of the detector is calculated using the density of the materials specified in the technical data sheets and the volume of each part, columns 5 and 6 respectively, as derived from its 3D mechanical design.

\begin{table} [!ht] 
    \centering
    \caption{Calculation of the actual mass of the miniBeBe detector according to the  mechanical concept. Al stands for aluminum and CF for Carbon fibre.}
  \begin{tabular}{m{30mm}|m{15mm}|m{30mm}|m{15mm}|m{20mm}|m{15mm}}
    \noalign{\hrule height 1pt}
        \hline
         Standard design & Mass (kg) & Optimised design & Mass (kg)& Volume per part (m$^{3}$) & Number of pieces  \\
          \noalign{\hrule height 1pt}
         \hline 
        *Electronic module  & 6.54 & *Electronic module & 6.54 & 8.54e-05 &  - \\
        Cooling plate & 1.75 &  Cooling plate & 1.75 & 5.742e-05 &  16\\
         Housing (CF)  & 14.65 & Housing (CF) & 14.65 & 3.84e-03 & 2 \\
         Flange (Al) & 1.92 & Flange (CF) & 1.37 & 7.18e-05 & 10 \\
         Locking ring (Al)  & 0.40 & Locking ring (CF) & 0.29 & 3.77e-05 & 4\\
         Support cables and pipes (Al) & 0.92 &  Support cables and pipes (CF) & 0.65 & 2.14e-05 & 16 \\
         \noalign{\hrule height 1pt}
         \hline
        Total mass& 26.18 & Total mass & 25.26\\
         \noalign{\hrule height 1pt}
         \hline
         * see Table \ref{Table_2} \\ 
         \noalign{\hrule height 1pt}
         \hline
    \end{tabular}
     \label{Table_7}
\end{table}

The analysis reveals that the replacement of materials for the fabrication of the flanges, fixing ring, and cable support results in a reduction in mass by approximately 4\%. A safety margin of 25\% was incorporated into the calculation of the sub-assembly mass to accommodate the electronic elements not included in the present analysis.

The largest contribution to the total detector mass comes from the housing. Therefore, it is preferable that carbon fibre, such as CFS-RI \cite{EC-TDS-High-Strength-Carbon-Sheet.pdf}, be considered as the replacement material. This possesses a low density, ranging from 1150 kg / m$^{3}$ to 1270 kg / m$^{3}$. This would result in a housing mass of 8.82 kg and a total detector mass of 15.65 kg, representing a reduction of 40.22\% compared to the currently proposed concept. Furthermore, the mass of the housing can be reduced by employing a material such as a carbon fibre sandwich comprising structural foam similar to Rohacell\textregistered \hspace {1pt} foam. It is anticipated that the overall reduction will be at least 45\%.

\subsection{Assembly sequence of miniBeBe detector version 2.0}

\begin{SCfigure}[0.8][!ht]
\caption{A flange compatible with the ITS mechanical housing is positioned on the mounting tool and adhesive is applied to the inside of the flange. The miniBeBe detector housing is then placed over these flanges, one at each end of the housing. The adhesive setting time is one week.} 
\includegraphics[width=0.6\textwidth]{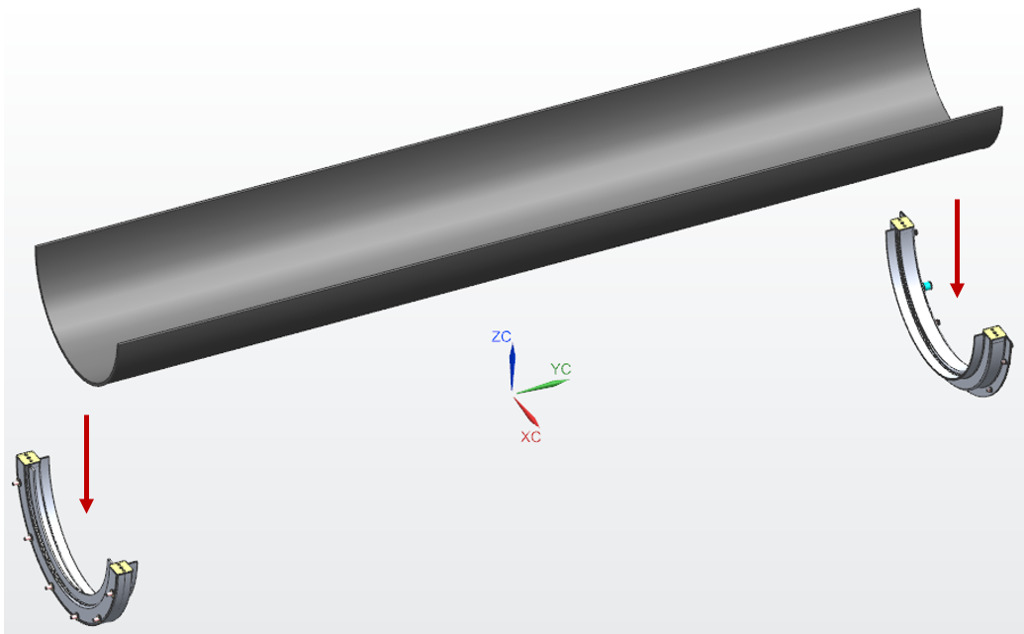}
\label{Figure_14} \\
\end{SCfigure}

\begin{SCfigure}[0.8][!ht]
\caption{The flanges of the miniBeBe detector and fixing rings are attached with adhesive and screws from the back of the housing. The distance between the flanges is 120 mm and the distance between the first and the last flange to the fastening ring is 145 mm.}
\includegraphics[width=0.6\textwidth]{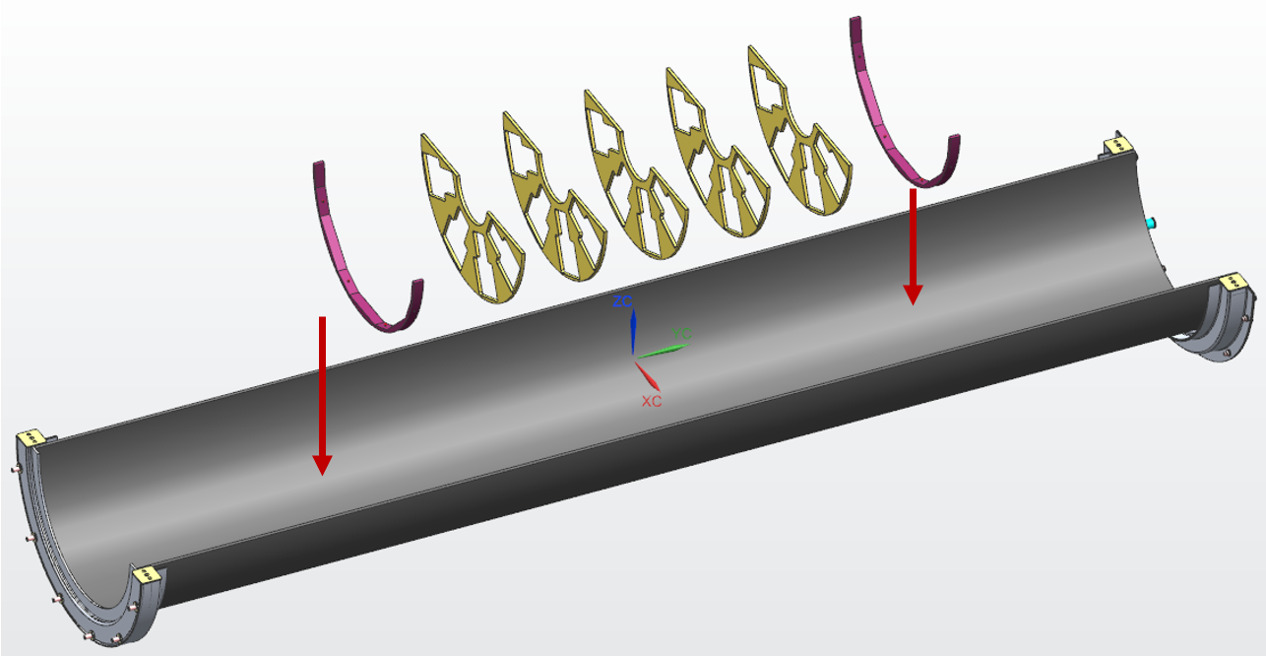}
\label{Figure_15}
\end{SCfigure}

\begin{SCfigure}[0.8][!ht]
\caption{Once the electronic modules are assembled, they enter through one of the ends and move across the windows of the flanges until they are placed over the locking ring. The ruby balls serve as guides to center each module. Then the module is screwed to the ring.}
\includegraphics[width=0.6\textwidth]{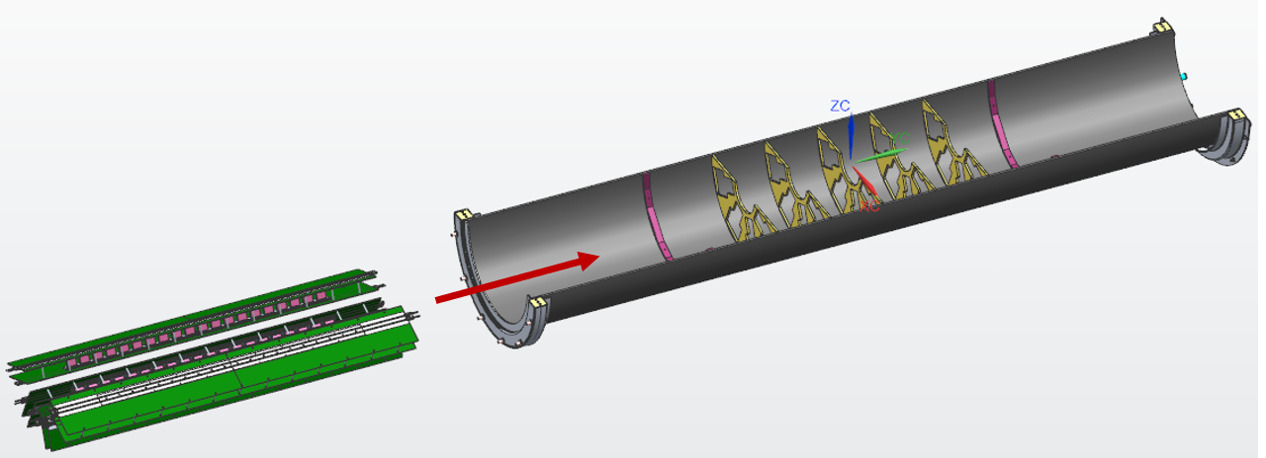}
\label{Figure_16}
\end{SCfigure}

\begin{SCfigure}[0.8][!ht]
\caption{The pipe and cable supports are fixed only with screws. In this manner, each module can be removed independently for replacement or maintenance. The distance from the ring to the support is 275 mm.}
\includegraphics[width=0.6\textwidth]{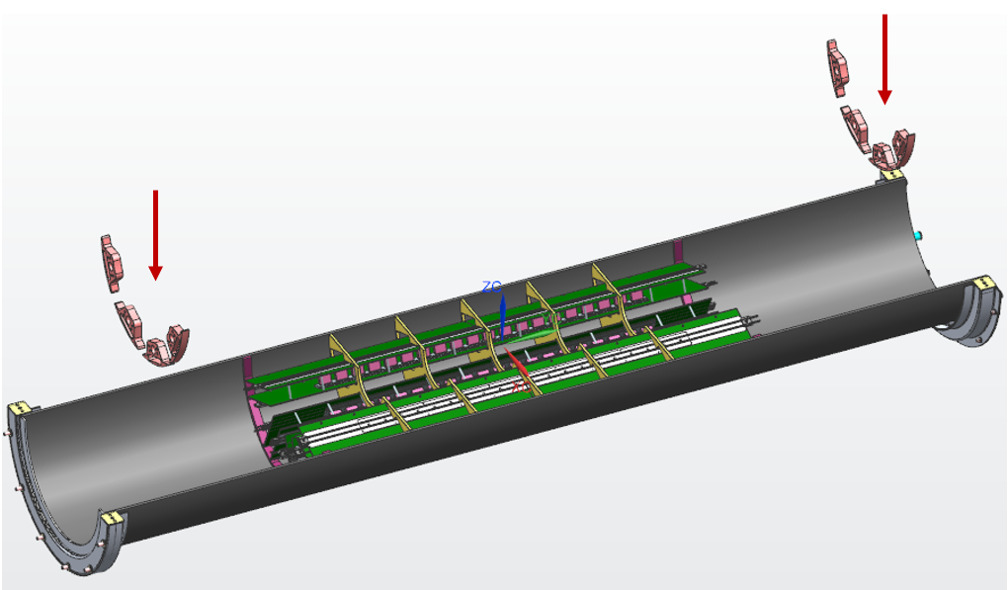}
\label{Figure_17}
\end{SCfigure}

\begin{SCfigure}[1][!ht]
\caption{After assembling the miniBeBe, the assembly tool is removed from the detector, which is then moved into its place within the MPD. Its position is the same as the one to be occupied by the ITS in the subsequent stage of the experiment.}
\includegraphics[width=0.55\textwidth]{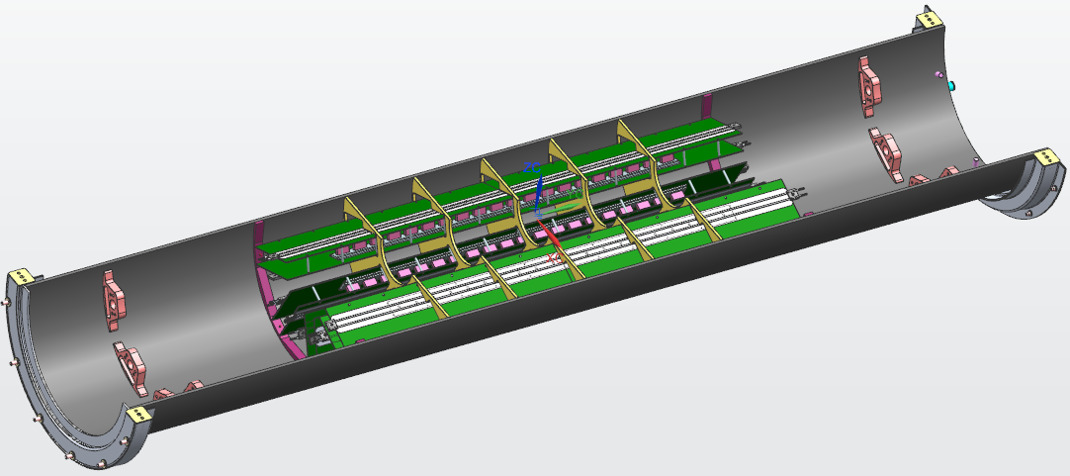}
\label{Figure_18}
\end{SCfigure}

\begin{SCfigure}[0.8][!ht]
\caption{Remove layers 1 to 4 of the ITS and then place the miniBeBe detector assembly, including the flange mentioned in point 1. The entire assembly will be placed on the flange of the fifth layer of the ITS.}
\includegraphics[width=0.6\textwidth]{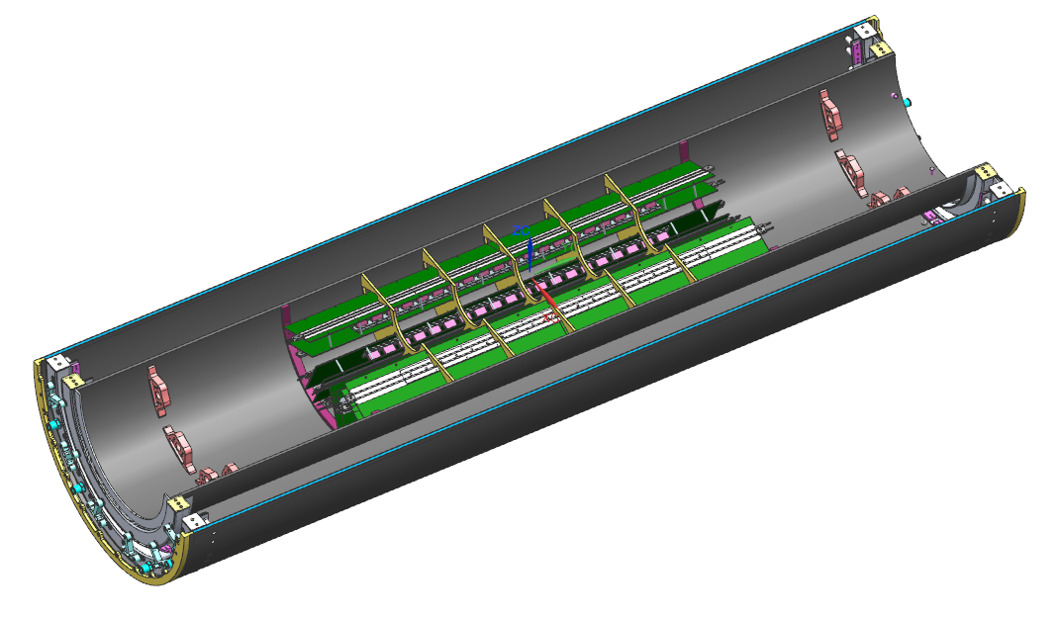}
\label{Figure_19}
\end{SCfigure}

To assemble all the components of the detector, a high-precision tool, previously manufactured for the ITS, is to be used. Araldite\textregistered \hspace{1pt} adhesive was used to secure the detector components to the detector housing with screws, and this adhesive is also used to assemble the electronics module-cooling plate sub-assembly. The sequence for assembling the detector is shown in Figs.~\ref{Figure_14} to~\ref{Figure_19}.

\section{Conclusions and future work}

%{\sout In this work, the development of the mechanical structure of the miniBeBe detector, conceived to contribute to the MPD TOF trigger for low multiplicity events, has been described.}
This work describes the development of the mechanical structure for the miniBeBe detector, designed to serve as a trigger for the Time of Flight Detector in the Multi-Purpose Detector (MPD) experiment for low multiplicity events.
The structure meets the requirements of not using ferromagnetic materials, as well as minimizing the material budget, while preserving a high resistance to deformation.

The design allows the detector to be placed inside the MPD, facilitating the removal of individual modules for maintenance or complete replacement after the experiment's first phase is completed. Removal of the miniBeBe detector will not affect the design of the other detectors in the experiment. Molds and additive manufacturing are part of the construction process. The maximum temperature on the cooling plate will be  $22.66^{\circ}$C and the outlet water temperature will be $21.41^{\circ}$C when the experiment enters into operation. The simulations show that the designed liquid cooling system will be able to stabilize the Silicon Photo-Multipliers (SiPMs) temperature, which are designed to be in contact with the plate. Recall that the working temperature of the SiPMs is within $-20.00^{\circ}$C to $80.00^{\circ}$C and thus the considered temperatures are well within the SiPMs operational range. In addition, a gas cooling system has been proposed which will help to reduce the temperature by about 1.00 to $2.00^{\circ}$C.

%The mass of each electronics module plus the support when assembled is 660 g. The total mass of the detector, when the cylinder container is fully assembled, is 21 kg, out of which 14 kg consist of the carbon fiber housing weight.
When assembled, each electronic module weighs 654.12 g. However, to be on the safe side, we added a 25\% mass for each sub-assembly to get 817.65 g., for a total weight of 6.54 kg for the eight modules. The carbon fibre housing of miniBeBe weighs 14.65 kg, so the total weight including all components should be 26.18 kg.

The deformation value of the detector is less than 1 mm with no shear stress, or von Mises stress, larger than 2 MPa. Therefore, there is no risk of mechanical failure due to fracture or detachment. We conclude that the design proposed in this work is suitable to be used as the mechanical structure of the miniBeBe detector.

Nevertheless, the design will likely suffer slight adjustments in the future, since it is conceived for version 1.0 of the electronics. It is expected that there will be at least two more versions before all the necessary electronic components are in place. A further challenge is to identify a suitable carbon fibre composite material, such as carbon foam and 3D printing with PLA-Carbon fibre, as well as a suitable manufacturing technique with the aim of replacing some aluminum parts and thus reducing the weight of the detector by at least $45{\%}$.

\acknowledgments

M. Herrera acknowledges support from a postdoctoral fellowship granted by CONAHCyT, Mexico. This work was supported in part by UNAM-DGAPA-PAPIIT grant number IG100322, CONAHCyT grant numbers CF-2023-G-433, A1-S-7655, and A1-S-16215 and CIC-UMSNH grant number 18371. The authors also thank for help and assistance received by Yu. A. Murin and other colleagues from the JINR LHEP STS department.

\bibliographystyle{JHEP}
\bibliography{References.bib}

\end{document}